\documentclass[12pt]{iopart}

\usepackage{hyperref}
\usepackage{graphicx}
\usepackage{epstopdf}
\usepackage{rotating}
\usepackage{array}
\usepackage{amssymb}
\usepackage{rotating}
\usepackage{slashed}
\usepackage{cite}

\def\v#1{{\bf#1}}
\def\be{\begin{equation}}
\def\ee{\end{equation}}
\def\bea{\begin{eqnarray}}
\def\eea{\end{eqnarray}}
\def\bfig{\begin{figure}}
\def\efig{\end{figure}}
\def\bcen{\begin{center}}
\def\ecen{\end{center}}
\def\bfl{\begin{flalign}}
\def\efl{\end{flalign}}

\def\ie{{\it i.e.\,}}

\def\<{\langle}
\def\>{\rangle}

\begin{document}

\title[On the electronic structure of benzene and borazine]{On the electronic structure of benzene and borazine: An algebraic description}

\author{Y.  Hern\'andez-Espinosa$^1$, R. A.  M\'endez-S\'anchez$^2$ and\\ E. Sadurn\'i$^3$}

\address{$^1$ Universidad Nacional Aut\'onoma de M\'exico, Instituto de F\'isica, A.P. 20-364, CDMX 01000, M\'exico}
\address{$^2$ Universidad Nacional Aut\'onoma de M\'exico, Instituto de Ciencias Fisicas, Apartado Postal 48-3, 62210 Cuernavaca Mor., M\'exico  }
\address{$^3$ Benem\'erita Universidad Aut\'onoma de Puebla, Instituto de F\'isica, Apartado Postal J-48, 72570 Puebla, M\'exico}
\ead{sadurni@ifuap.buap.mx}

\begin{abstract}
The spectrum of a hexagonal ring is analysed using concepts of group theory and a tight-binding model with first, second and third neighbours. The two doublets in the spectrum are explained with the $C_3$ symmetry group together with time-reversal symmetry. Degeneracy lifts are induced by means of various mechanisms. Conjugation symmetry breaking is introduced via magnetic fields, while $C_3$ breaking is studied with the introduction of defects, similar to the inclusion of fluorine atoms.  Concrete applications to benzene and borazine are shown to illustrate the generality of our description. Wave functions are described in connection with partial or full aromaticity. Electronic density currents are found for all cases. A detailed study of a supersymmetry in a $6$-ring is presented and its consequences on electronic spectra are discussed.
\end{abstract}

\maketitle

\section{Introduction}
The electronic structure of benzene, understood in terms of valence orbitals, has been known to chemists for a long time \cite{cooper_electronic_1986} and even exposed in anecdotal fashion in \cite{benfey_august_1958}. Quite recently, a surge on this subject involving the more elusive borazine \cite{armstrong_electronic_1970} --also known as the inorganic benzene-- shows a pursuit for the best of descriptions regarding its electronic structure. For example \cite{Parker_2018}, the vibrational levels of borazine were obtained experimentally with neutron scattering techniques, together with DFT calculations in bulk. The aromatic structure of these rings has been discussed for some time \cite{Kiran_2001} with increasingly detailed techniques \cite{cooper_electronic_1986}. While delocalization in benzene is now incontrovertible when degeneracy is correctly considered, the case of borazine seems to be more consistent with partial localization or moderate aromaticity. DFT was also employed for borazine \cite{Shen_2007} supporting this view. It should be mentioned that long-standing experiments regarding the atomic and electronic configurations of these molecules have provided a solid basis for theoretical discussions: The lower part of the electronic spectrum in benzene was studied already in \cite{Clark_1967} with photoelectron spectroscopy, and electron impact experiments were reported in \cite{Doering_1977}. The older calculations by Moskowitz \cite{Moskowitz_1963} already suggested the use of symmetry-adapted bases as well as $\pi-\sigma$ interactions between orbitals, in order to understand the resulting electronic clouds. As a conclusion, the groups $C_6$ and $C_3$ and their adapted functions should never be disregarded. 

Our goal in this work is to provide a simple algebraic description of energy levels and wave functions that reasonably reproduce the known properties of 6 rings. We shall work with group theoretical concepts that include the representations of cyclic groups $C_3$, $C_6$, $Z_2$  and the introduction of antiunitary operators in connection with magnetic perturbations. Also, a supersymmetry $N=2$ related to a $Z_2$--grading \cite{Kac_1977, Haag_1975} will make its appearence in a six-dimensional Hilbert space. Our treatment will be closely connected to specific matrix realizations and shall keep calculations as explicit as possible, for the sake of clarity. We should also mention that, despite the extensive use of Galois fields in highly symmetric situations, their correct composition in phase space was only given recently in \cite{Wootters_1987, Vourdas_2006, Vourdas_2005, Vourdas_1993, Zak_2005} for odd cases and \cite{Sadurni_2019} for even cases, which are of particular importance for 6 rings as the group decomposition $C_6=C_3 \times Z_2$ trivially shows.

As a further motivation of our work, the simplicity of the present approach suggests the use of simple molecular orbitals for other computational applications. Our recent incursion into hexagonal structures \cite{sadurni_diracmoshinsky_2011,Sadurni_2019} sprouted some interest on exotic effects emerging even in the simplest of descriptions: tight-binding models of a single electron. In the realm of quantum emulations, realizations in microwaves \cite{stegmann_vortices_2019, Kuhl_2010,Barkhofen_2013,Villafane_2013,Bellec_2013}, and elastic waves \cite{Torrent_2012,Torrent_2013,Zhong_2011}  have shown the universality of frequency spectra produced by highly symmetric structures, such as hexagon and triangle symmetries, together with their moderate breaking.

Structure of the paper: section \ref{sec:1} presents the tight-binding ring model benzene-like molecules. The energy levels are obtained analytically and the presence of two doublets are identified with time-reversal symmetry. A discrete version of the continuity equation is studied and expressions for electronic currents revolving around the molecule are obtained. Numerical calculations for benzene and borazine are developed using reported parameters in the literature. The borazine model is studied in section \ref{sec:2}. The Hamiltonian is presented as a partitioned $2\times2$ diagonal matrix where each block is solvable by radicals. Finally, in section \ref{sec:3}, a dynamical supersymmetry (SUSY)  in the ring is found and presented. The dynamical SUSY provides a theoretical explanation of the spectral symmetry under level reflection around the an isolated carbon atom energy. This holds for both benzene and borazine spectra, even in a globally broken $C_3$ configurational symmetry. Conclusions are given in \ref{sec:4}.


\section{Benzene and the $C_6$ group\label{sec:1}}
The molecular structure of benzene was a subject of study since the second half of the XIX century, when the chemist August Kekul\'e recognized that there should exist a symmetry inherent to the molecule and elucidated the hexagonal geometry of benzene.
With the development of the quantum theory, the atomic structure of benzene became clearer \cite{pauling_shared-electron_1928,empedocles_electronic_1964}. Each of the six carbons in the molecule possesses three $sp^2$ hybrid orbitals, which are coordinated with other two $C$ atoms and with the $s$ orbital of the  $H$ atom, leading to $\sigma$ states. The six $p_z$ atomic orbitals of the $C$ atoms perpendicular to the molecular plane, form $\pi$ bonds; the six electrons corresponding to these orbitals are delocalized when found in stationary states as they do not belong to any particular $C$ centre. The borazine model can be constructed in a similar way to benzene, where the orbital hybridization $sp^2$ is proposed to explain the molecular structure, despite the fact that internal angles differ from those of a perfect hexagonal geometry \cite{Gauss_2000}. The role of symmetry and algebraic methods should never be underestimated, as their neglect might lead to unusual predictions regarding the localization of electrons around atomic centres \cite{cooper_electronic_1986} even after diagonalization takes place.

In this section a tight-binding model is presented using a hexagonal ring with localized wave functions around each site. Although the problem of effective orbitals immersed in structures has not been solved in chemistry, i.e. no analogues of Wannier functions are known in this area, it will be reasonable to assume that atomic orbitals can be employed as examples of such highly localized waves.

On quantitative grounds, we justify our approximations by employing on-site energies of C atoms of $-8.97$ eV \cite{Harrison_1989}, nearest couplings of $2.7$ eV for C-C  at a distance of $1.4$ \AA \cite{Zhao_2010} consistent with the band structure of graphene and its Fermi velocity \cite{Geim_2007}, nearest couplings of $1.95$ eV between N-B that are consistent with the electronic bands of boron nitride, and exponential decay of hopping integrals with the distance between atomic centres. It should be noted that experimental values of the energy distance between a singlet and a doublet in benzene (both $\pi$ orbitals) were reported \cite{Lindholm_1967} as $(12.1- 9.3)$ eV $= 2.8$ eV which are also consistent with our results, as long as the on-site energy is also corrected by adding between $5$ eV and $6$ eV accounting for Coulomb repulsion of electron charges at a distance of $1.4$ \AA. A 'back of the envelope' estimate for a hollow charged sphere yields $(1/2) \times 10$ eV, while a solid ball of the same radius has an energy of $(3/5) \times 10$ eV. For a ring, we expect a similar geometric factor. With our simplified set of parameters, we may also introduce a defect, simulating the presence of a flourine atom, as considered by \cite{Clark_1967} and \cite{Tegeler_1980}.

\subsection{Multiple neighbour tight-binding model}
The first system of study is a $6$-ring where all the sites are equal; this corresponds to the benzene case. We start with a multiple-neighbour model and write the Hamiltonian $H$ in matrix representation as:
\be
\label{eq1}
H\doteq
\left(
\begin{array}{ccc|ccc}
\epsilon&\Delta_{12}&\Delta_{13}&\Delta_{14}&\Delta_{15}&\Delta_{16}\\
\Delta^*_{12}&\epsilon&\Delta_{23}&\Delta_{24}&\Delta_{25}&\Delta_{26}\\
\Delta^*_{13}&\Delta^*_{23}&\epsilon&\Delta_{34}&\Delta_{35}&\Delta_{36}\\
\hline
\Delta^*_{14}&\Delta^*_{24}&\Delta^*_{34}&\epsilon&\Delta_{45}&\Delta_{46}\\
\Delta^*_{15}&\Delta^*_{25}&\Delta^*_{35}&\Delta^*_{45}&\epsilon&\Delta_{56}\\
\Delta^*_{16}&\Delta^*_{26}&\Delta^*_{36}&\Delta^*_{46}&\Delta^*_{56}&\epsilon
\end{array}
\right),
\ee
where $\Delta_{ij}=\Delta^*_{ji}$ in hermitian models (electrons are neither gained nor lost from the system, so probability is conserved). The Schr\"odinger equation possesses time-reversal symmetry if all couplings $\Delta_{ij}$ are real. Since overlaps are given by
\be\label{eq2}
\Delta_{ij}=\int \Psi^*_i(\mathbf{x}) H \Psi_j(\mathbf{x})d^3 x,
\ee
this means that a Hamiltonian $H= (p^2/2m)+V(\mathbf{x})$ for a single electron without magnetic fields contains $\Delta_{ij}\in \mathbb{R}$ if and only if $\Psi_i \in \mathbb{R}$. This is indeed the case if the function $\Psi_i(\mathbf{x})=\<\mathbf{x}|i\>$, ($|i\>$ localized around site $i$) corresponds to a bound state of an isolated carbon atom, or in general, a state without outgoing probability current.\footnote{Wannier functions are commonly used in solid state physics, but to our knowledge the have not been defined in molecular settings; this would entail the use of finite Fourier transforms}

\bfig
\bcen
\includegraphics[width=15cm]{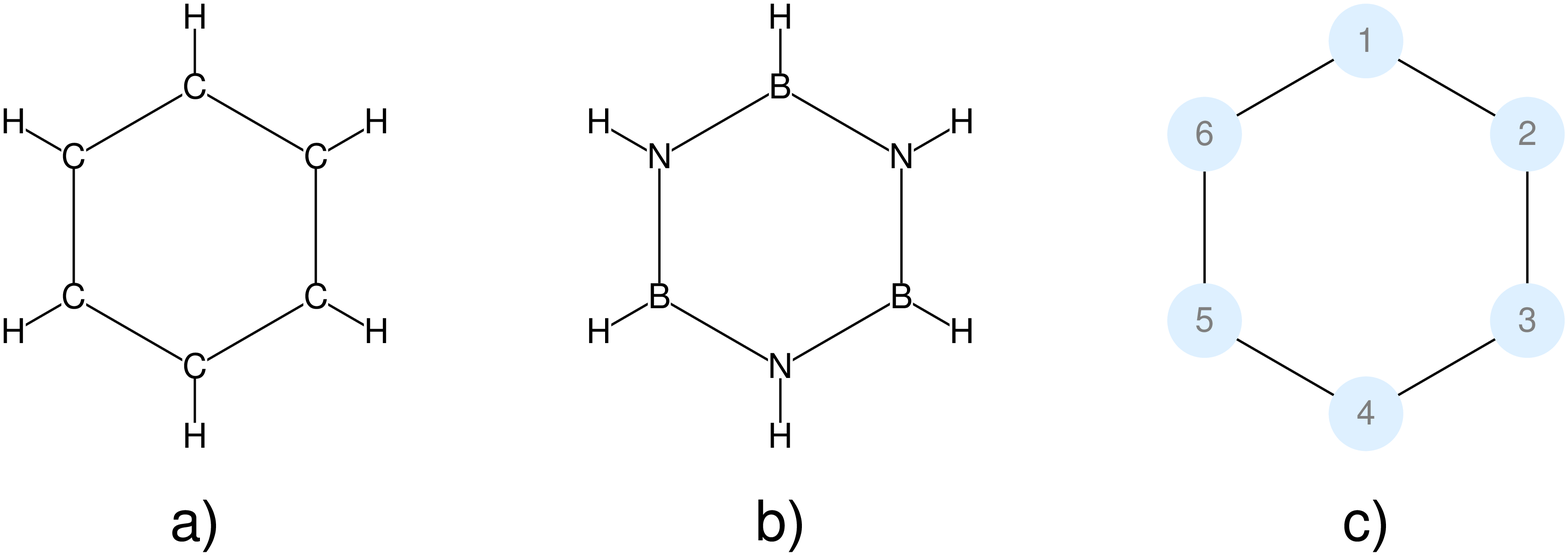}
\caption{\label{fig1}a) Molecular structure of benzene. Each carbon is coordinate with two carbons and one hydrogen atom.  b) Molecular structure of borazine. The vertices of the hexagon have boron and nitrogen atoms alternated.  c) Six-site model for benzene. The hydrogen uses its only available atomic state to establish the covalent bonding with the carbon atom. On the other hand the carbon uses a $sp^2$ hybridization to form the necessary chemical bondings for the benzene molecule. The only atomic orbital that contributes to the electronic conduction is the $p_z$ which is perpendicular to the molecular plane. }
\ecen
\efig

Since our first approach to the problem corresponds to a highly symmetric configuration (such as a regular hexagon) we start by constraining the couplings $\Delta_{ij}$. A translation operator modulo $6$ consists of a rotation around the hexagon centre by $\pi/3$ radians. This operation is represented by $\hat{T}|i\>=|(i+1)\,\mbox{\small mod}\, 6\>$, $\hat{T}\hat{T}^\dagger = \hat{T}^\dagger \hat{T} = \mathbf{1}$, $[\hat{T},\hat{T}^\dagger]=0$ and with $\{ |i\> \}$ the canonical basis, we have
\be
\hat{T}
\doteq
\left(
\begin{array}{cccccc}
0&0&0&0&0&1\\
1&0&0&0&0&0\\
0&1&0&0&0&0\\
0&0&1&0&0&0\\
0&0&0&1&0&0\\
0&0&0&0&1&0
\end{array}
\right).
\ee
For $\hat{T}$ to be a symmetry of $H$, one must have $[H,\hat{T}^q]=0$ for all $q=1,...,6$. The Abelian group $C_6$ then arises $C_6=\{\mathbf{1},\hat{T},\hat{T}^2,\hat{T}^3,\hat{T}^4,\hat{T}^5\}$. Also, the space inversion $1 \mapsto 1$, $2\mapsto 6$, $3 \mapsto 5$, $4 \mapsto 4$, $5 \mapsto 3$, $6 \mapsto 2$ is a symmetry of $H$, but it does not commute with $\hat{T}$. So now all secondary diagonals in (\ref{eq1}) must have equal elements, leading to the notation
\be
H\doteq
\left(
\begin{array}{cccccc}
\epsilon&\Delta_1&\Delta_2&\Delta_3&\Delta_2&\Delta_1\\
\Delta_1&\epsilon&\Delta_1&\Delta_2&\Delta_3&\Delta_2\\
\Delta_2&\Delta_1&\epsilon&\Delta_1&\Delta_2&\Delta_3\\
\Delta_3&\Delta_2&\Delta_1&\epsilon&\Delta_1&\Delta_2\\
\Delta_2&\Delta_3&\Delta_2&\Delta_1&\epsilon&\Delta_1\\
\Delta_1&\Delta_2&\Delta_3&\Delta_2&\Delta_1&\epsilon
\end{array}
\right),
\ee
where the subscript in $\Delta$ now denotes the neighbouring sites; obviously there is no $\Delta_4$. In this notation
\be\label{eq5}
H=\epsilon+\sum_{n=1}^3 \left(\Delta_n \hat{T}^n + \Delta_n(\hat{T}^\dagger)^n\right),
\ee
as the reader may readily verify by using the explicit powers of $\hat{T}$. The following matrix relations shall be useful

\bfig
\bcen
\includegraphics[width=15cm]{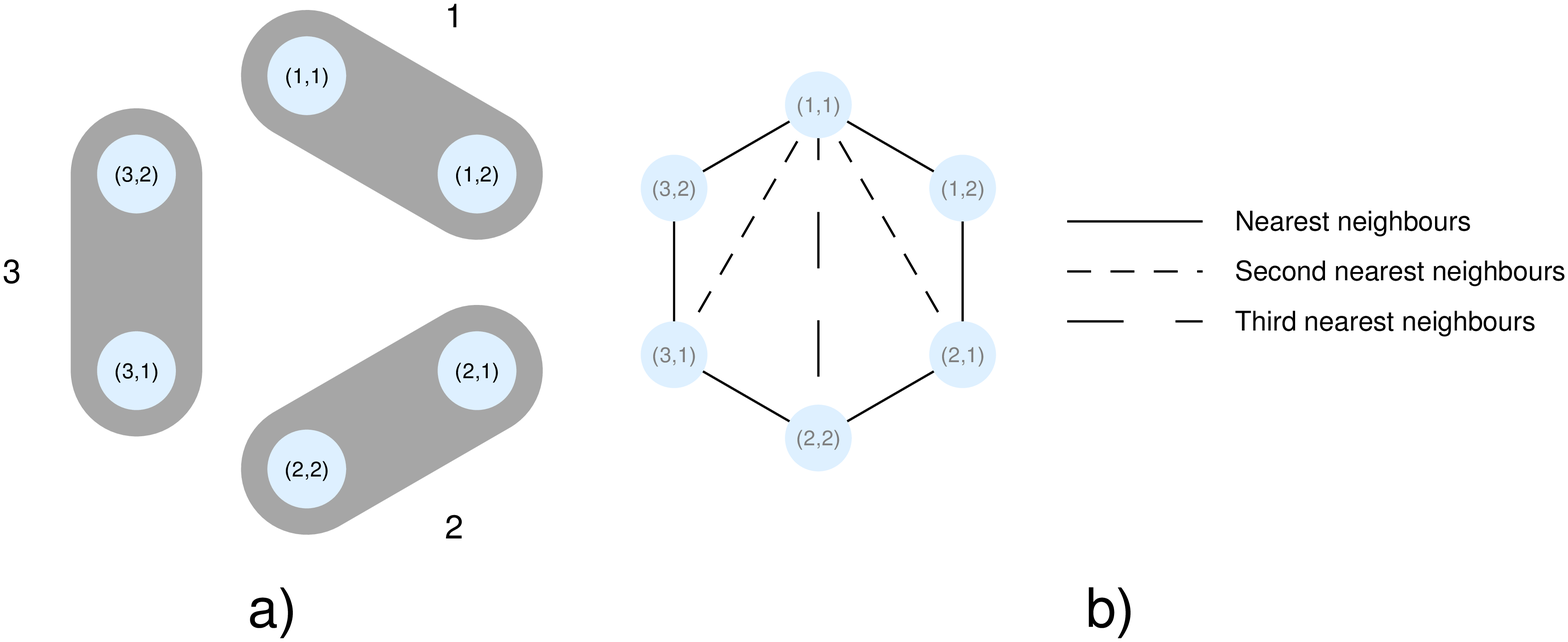}
\caption{\label{fig2} a) Decomposition of the hexagon structure into three dimers. The sites are labelled with a pair of numbers $(n,m)$ where $n$ stands for the dimer number and $m$ for the internal position in one dimer. b) Graphical representation of nearest, second and third neighbour couplings in a hexagonal ring.  }
\ecen
\efig

\bea
\hat{T}^3 =
\left(
\begin{array}{cccccc}
0&0&0&1&0&0\\
0&0&0&0&1&0\\
0&0&0&0&0&1\\
1&0&0&0&0&0\\
0&1&0&0&0&0\\
0&0&1&0&0&0
\end{array}
\right)=(\hat{T}^\dagger)^3, \\ \nonumber
\hat{T}^2=
\left(
\begin{array}{cccccc}
0&0&0&0&1&0\\
0&0&0&0&0&1\\
1&0&0&0&0&0\\
0&1&0&0&0&0\\
0&0&1&0&0&0\\
0&0&0&1&0&0
\end{array}
\right)=\hat{T}^4=(\hat{T}^\dagger)^{-4}=(\hat{T}^\dagger)^{-2},\\ \nonumber
\hat{T}^5=\hat{T}^{-1}=\hat{T}^\dagger .
\eea
With these simple expressions we are able now to diagonalize $H$, as it only depends on $\hat{T}$, $\hat{T}^\dagger$. It suffices to find the eigenbasis of $\hat{T}$. We have $\{|q\>\}, \, q=1,...,6$ such that $\hat{T}|q\>=\exp(i2 \pi q/6)|q\>$ where the exponential comes from the unitarity of $\hat{T}$ and the form of the exponent is obtained from the fact that $\hat{T}^{6n}=\mathbf{1}$ for all $n$, so $\exp(i2 \pi q (6n)/6)=\exp(i2 \pi qn)=1$. We have now eigenvectors such that
\bea
|q\>=\sum_{i=1}^6 C_i^q |i\> \quad \mbox{\small and}\quad \hat{T}|q\>=\sum_{i=1}^6 C_i^q \hat{T}|i\>=\sum_{i=1}^6 C_i^q|(i+1)\, \mbox{\small mod}\, 6\>, \nonumber \\
\hat{T}|q\>=\sum_{i=1}^6 C_{(i-1)\, \mbox{\scriptsize mod}\, 6}^q |i\>=\sum_{i=1}^6 e^{i2\pi q/6}C_i^q|i\>.
\eea
Since $C_i^q = \exp(-i\pi q/3)C_{(i-1)\, \mbox{\scriptsize mod}\, 6}^q$ and the normalization is $1/\sqrt{6}$, the coefficient reads
\be\label{eq8}
C_j^q = e^{-i\pi qj/3}/\sqrt{6}\quad i=1,...,6.
\ee

The explicit coefficients (\ref{eq8}) together with $|q\>$ above constitute already the eigenfunctions of the problem. 
Now we apply $H$ to our expression for $|q\>$ and find easily that
\be
H|q\>=\left\{\epsilon + \sum_{n=1}^3\left(\Delta_n e^{in\pi q/3}+\Delta_n e^{-in\pi q/3 } \right)  \right\}|q\>,
\ee
thus the energy becomes
\be
\label{eq9}
E_q= \epsilon+2\sum_{n=1}^3 \Delta_n\cos \left(\frac{\pi q n}{3} \right).
\ee
This shows that for multiple neighbours with arbitrary couplings, the spectrum has certain properties inherent to the symmetry, and not to the specific details of the overlaps $\Delta_n$ related to tunnelling. Therefore the explicit expression for tunnelling rates between sites is at this point irrelevant. For instance, we have that if $q \mapsto -q$ then $E_q =E_{-q}$. But $-q=(6-q) \, \mbox{\small mod} \,6$ according to (\ref{eq9}), therefore the following pairs are degenerate $q=1,5$; $q=2,4$ while $q=3$ and $q=6$ are singlets. We show a plot in figure \ref{fig4} panel a).

\subsection{Magnetic field piercing molecular planes}
Now we introduce a magnetic field orthogonal to the molecular plane. Going back to our single-electron Hamiltonian
\be
H=\frac{\left[\mathbf{p}- (e/c)\mathbf{A} \right]^2}{2m}+V(\mathbf{x})\neq H^*
\ee
due to the presence of terms $(e/2mc)\mathbf{p}\cdot \mathbf{A}$ and $(e/2mc)\mathbf{A}\cdot \mathbf{p}$ when the square is expanded (evidently $\mathbf{p}=(-i\hbar \vec{\nabla})^*=i\hbar \vec{\nabla}=-\mathbf{p}$).

From here and the tight binding representation of $H$ limited to a single carbon valence orbital (our $6\times 6$ matrix) we find that $\Delta_{ij}$ cannot be real any more. To see the relation with a magnetic field, we must apply a unitary transformation that removes all the phase factors of $\Delta_{ij}$ in as much as possible. If the transformation
\be
U_{\mbox{\scriptsize gauge}} \doteq \mbox{diag}\left\{e^{i\Phi_1},...,e^{i\Phi_6} \right\}, \quad H'=U_{\mbox{\scriptsize gauge}}^\dagger H U_{\mbox{\scriptsize gauge}}
\ee
removes all phases, then there cannot be a magnetic flux through the molecule, and no overall effect can be found in the spectrum. (In passing, we note that for open chains this is always the case, so the presence of closed loops in the coupling diagram is essential). On the contrary, if a phase-factor always remains after $U_{\mbox{\scriptsize gauge}}$ is applied, then we may have at least a certain element $\Delta_n e^{i\Phi_n}$ where $\Phi_n$ has accumulated all phases in $H$, to give
\be
H'=
\left(
\begin{array}{ccc|ccc}
\epsilon&\Delta_1&\Delta_2&\Delta_3&\Delta_2 e^{i\Phi_2}&\Delta_1 e^{i\Phi_1}\\
\Delta_1&\epsilon&\Delta_1&\Delta_2&\Delta_3&\Delta_2 e^{i\Phi_2}\\
\Delta_2&\Delta_1&\epsilon&\Delta_1&\Delta_2&\Delta_3\\
\hline
\Delta_3&\Delta_2&\Delta_1&\epsilon&\Delta_1&\Delta_2\\
\Delta_2 e^{-i\Phi_2}&\Delta_3&\Delta_2&\Delta_1&\epsilon&\Delta_1\\
\Delta_1 e^{-i\Phi_1}&\Delta_2 e^{-i\Phi_2}&\Delta_3&\Delta_2&\Delta_1&\epsilon
\end{array}
\right).
\ee
However, for notational simplicity, we write the Hamiltonian $H$ in a gauge such that
\be
\Delta_n = e^{i\phi _n}|\Delta_n|
\ee
for all couplings. With this, the new Hamiltonian becomes
\be
H=\epsilon + \sum_{n=1}^3\left( e^{i\phi_n}|\Delta_n|\hat{T}^n + e^{-i\phi_n}|\Delta_n|(T^\dagger)^n\right).
\ee
It is remarkable that $|q\>$ still solves the eigenvalue problem in this case, producing thus
\be
E_q (\{\phi_i\})=\epsilon +2\sum_{n=1}^3 |\Delta_n|\cos \left(\frac{\pi q n}{3}+\phi_n \right),
\ee
where $E_q (\{\phi_i\})$ is the flux-dependent energy. As the reader may see, the property $q\mapsto -q$ is now broken:
\bea
E_{-q}(\{\phi_i\})&=\epsilon + 2\sum_{n=1}^3|\Delta_n|\cos \left(\frac{-\pi q n}{3}+\phi_n \right)\\
&=\epsilon +2 \sum_{n=1}^3 |\Delta_n|\cos \left( -\phi_n + \frac{\pi q n}{3}\right)\\
&=\epsilon + 2\sum_{n=1}^3 |\Delta_n|\left\{\cos\left(\frac{\pi q n}{3} \right)\cos \phi_n + \sin\left(\frac{\pi q n}{3} \right)\sin\phi_n \right\}.
\eea
The energy gap between $E_q$ and $E_{-q}$ is
\be
E_{-q}-E_q=\Delta E=4\sum_{n=1}^3|\Delta_n|\sin\phi_n\sin\left(\frac{\pi q n}{3} \right)
\ee
and now we see that $\sin (\pi q n/3)=-\sin (-\pi q n/3)$ produces a difference. The conclusion is that $E_{-q} (\{\phi_i \})$ and $E_{q}(\{\phi_i \})$ are no longer degenerate and the two doublets are split by an amount controlled by the field $\phi_n$. For example, if a nearest-neighbour model is considered, \ie  $\Delta_1\equiv \Delta$, $\Delta_2=\Delta_3=0$ then a rotation of the phases expresses the level repulsion quantitatively as shown in figure \ref{fig4} panel d).

\bfig
\bcen
\includegraphics[width=12cm]{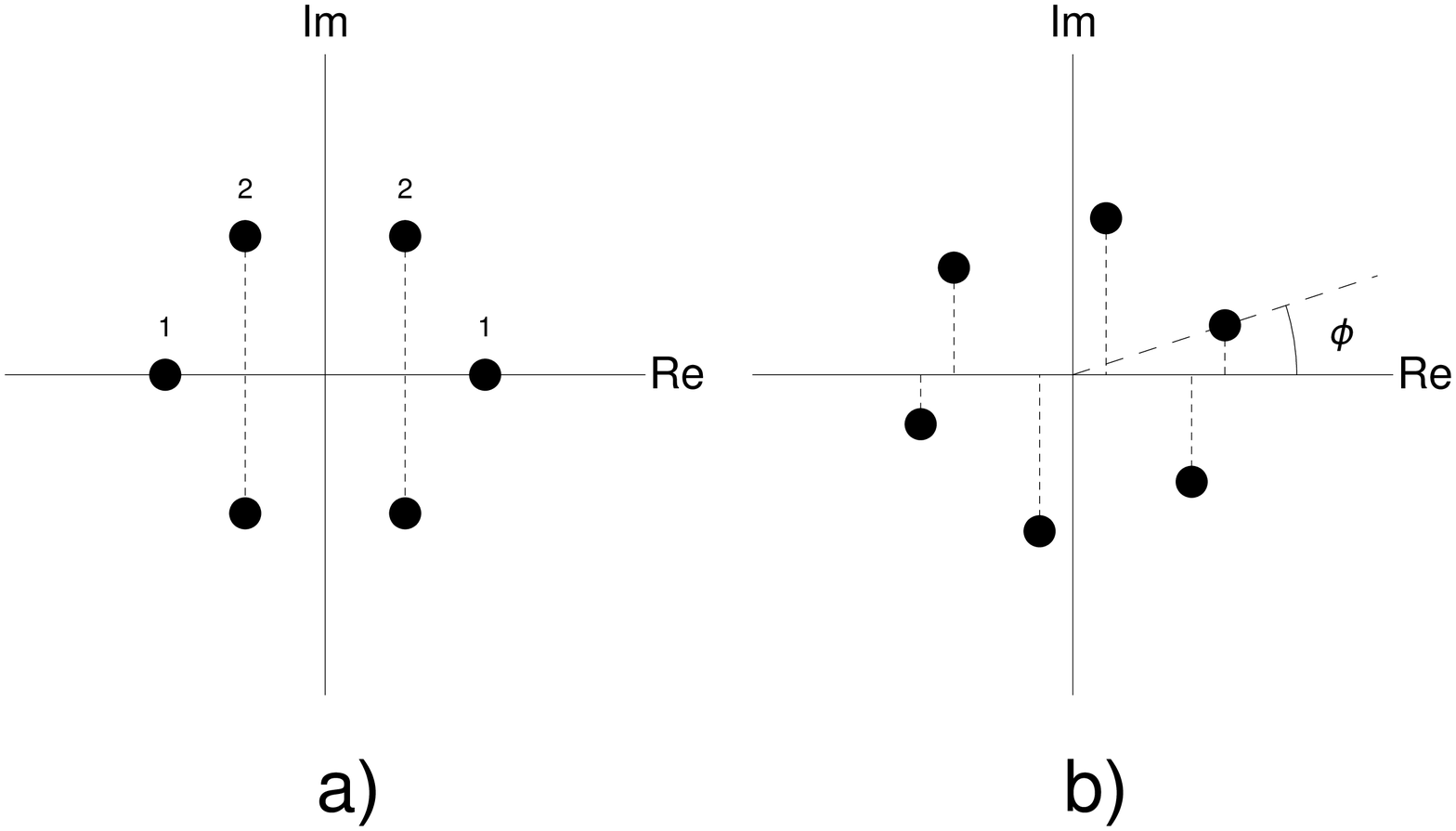}
\caption{\label{fig3} a) Eigenphases of the $C_6$ group. The level structure can be inferred from this diagram, as the eigenphases with the same real projection correspond to degenerate states. In this way a spectrum with singlet, doublet, doublet and singlet can be visualized. b) In the presence of a magnetic field an extra phase is introduced and the corresponding projections of the degenerate states on the real axis are not the same as in a),  so lifted doublets appear.  }
\ecen
\efig

\subsection{Other symmetry breaking}

For certain models we have that $E_q-\epsilon=\epsilon-E_{q'}$ \ie for some pairs $E_q+E_{q'}=2\epsilon$ holds. This implies that
\be
\sum_{n=1}^3|\Delta_n|\left\{\cos\left(\phi_n + \frac{\pi q n}{3} \right)+\cos\left(\phi_n + \frac{\pi q' n}{3} \right) \right\}=0.
\ee
Let $q=1$ and $q'=6$ (or $q=2$, $q'=5$; $q=3$, $q'=4$). We then have $q'+q=1\, \mbox{\small mod}\, 6$. From here it follows that
\bea
\sum_{n=1}^3|\Delta_n|\left\{\cos\left(\frac{(1-q)\pi n}{3}+\phi_n \right)+\cos\left(\frac{q\pi n}{3}+\phi_n \right) \right\}=0
\eea
for all $\Delta_n$, so equating the cosine arguments up to a difference of $\pi$ leads to  $(q\pi n/3)=(\pi n/3)-(q\pi n/3)\pm \pi$ or $2q n =n\pm 3$ for all q. Three cases arise depending on the number of neighbours entering the tight-binding model:
\begin{itemize}
\item If $n=1$ (nearest neighbours), $2q\pi=-2\pi \, \mbox{\small mod}\, 2\pi$ or $2q\pi=4\pi$, valid for all $q$.
\item If $n=2$ (second neighbours), $4q\pi\neq 5\pi \, \mbox{\small mod}\, 2\pi$ and there is no solution for $q$.
\item If $n=3$, $2q\pi=0 \, \mbox{\small mod}\, 2\pi$ for all $q$ and for both signs $\pm \pi$ considered above.
\end{itemize}
It is thus concluded from the second case that only second neighbours break the reflection symmetry of the spectrum around $\epsilon$ even when magnetic flux is introduced, while first and third neighbours preserve it.

\bfig
\bcen
\includegraphics[width=12cm]{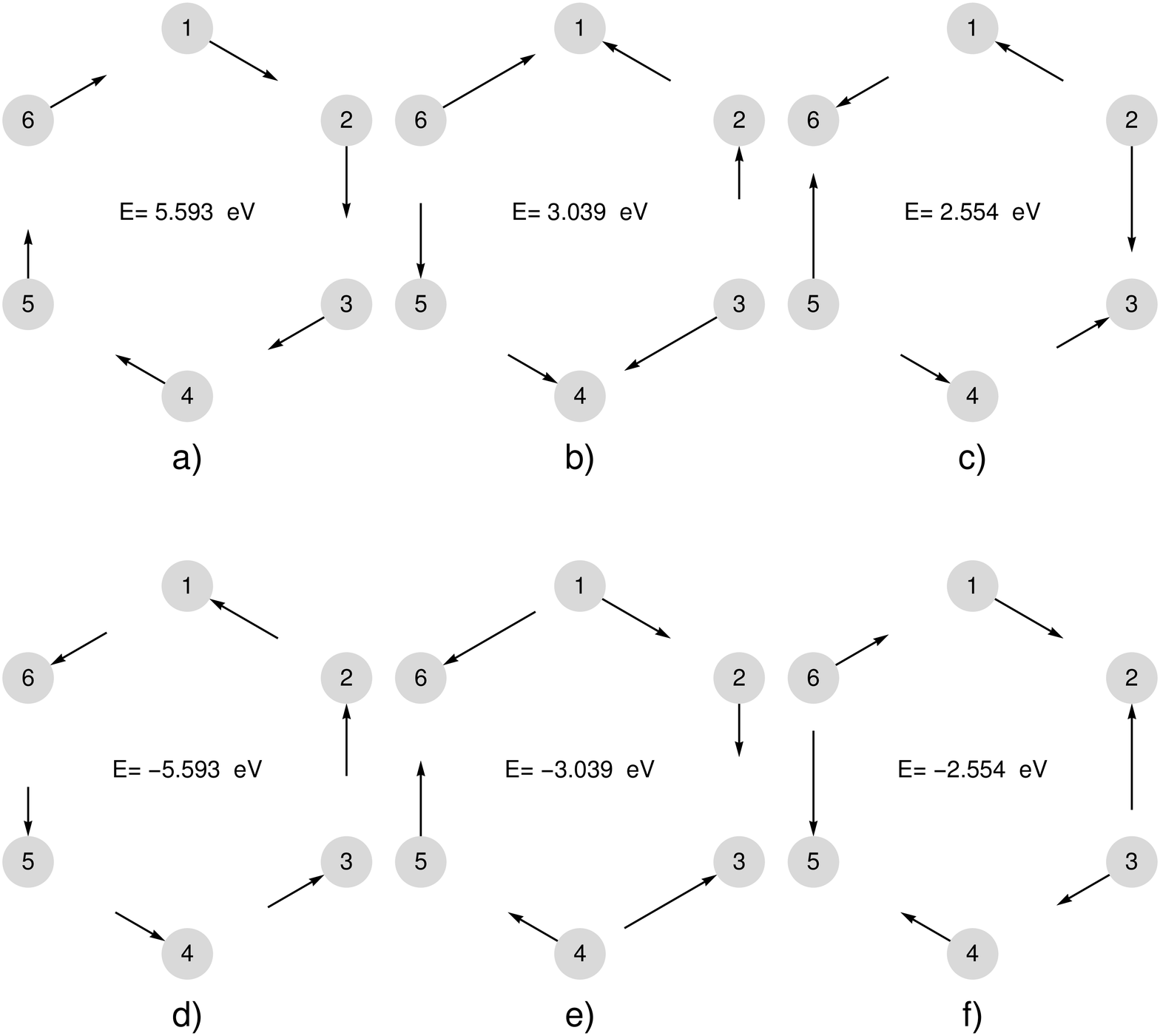}
\caption{\label{fig4.1} Current plots for the six eigenfunctions of the benzene molecule under the action of a magnetic field of $B=8.5855 \times 10^{-3}$G  perpendicular to the ring plane. Please note that the superior and inferior rows correspond to reflected energy levels, the only way to distinguish between them is the current scheme.
 Panels a) and d) correspond to singlet states. Panels b), c) and e), f) show the inequivalent current schemes for the superior and inferior lifted doublets respectively. It is worth noting that both rows are inverted current schemes of each other.}
\ecen
\efig

\bfig
\bcen
\includegraphics[width=12cm]{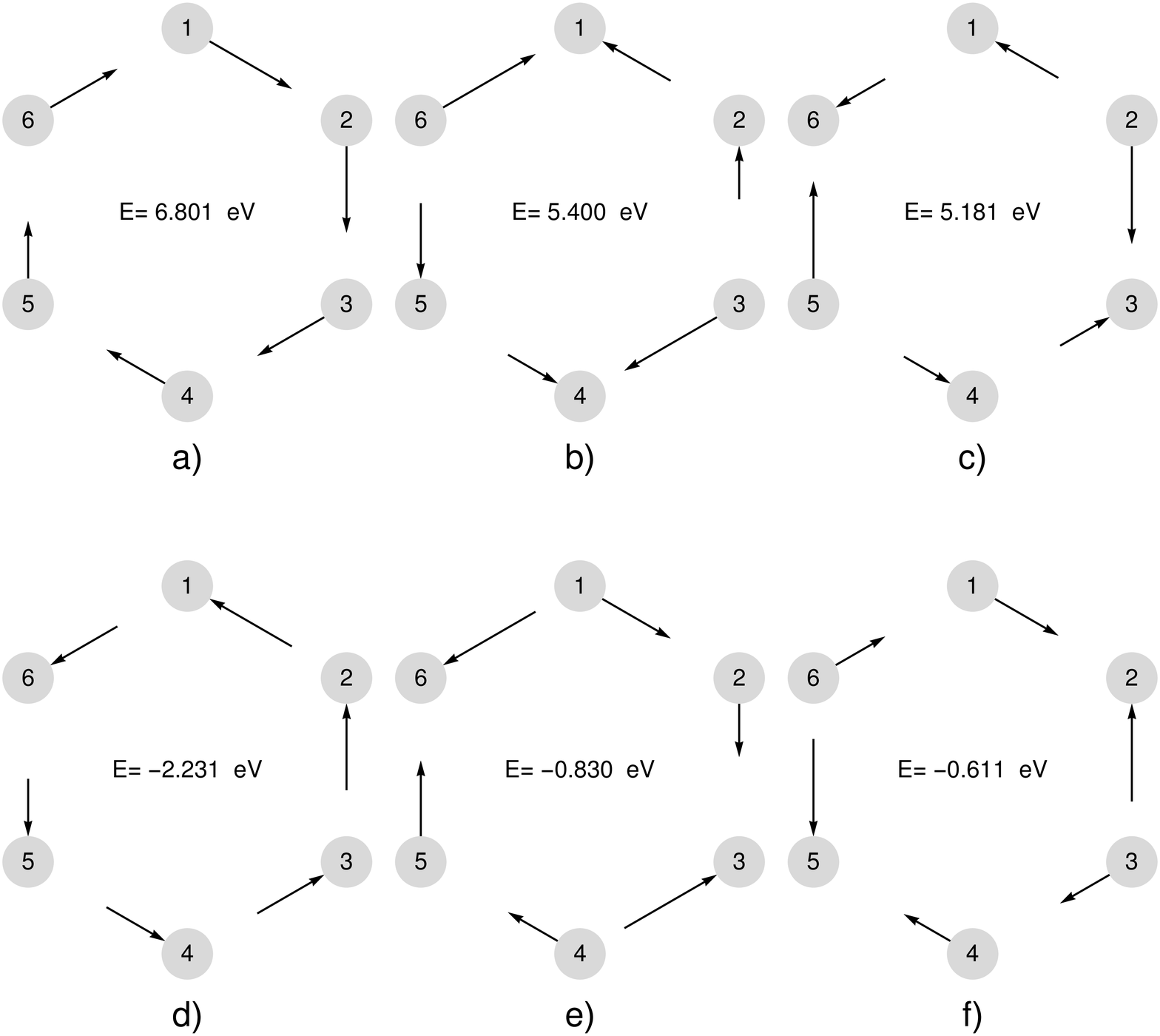}
\caption{\label{fig4.2} Current plots for the eigenfunctions of the borazine molecule under a magnetic field of $B=8.5855 \times 10^{-3} G$. Similarly to the benzene case, the singlets of panels a) and d) show opposite current directions. Panels b), c) and e), f) correspond to the lifted doublets states. The current schemes are different and make a distinction between reflected energy levels.  }
\ecen
\efig

\subsection{Eigenfunctions and discrete currents}


From the definition of current using our single electron Hamiltonian, we write
\be
\mathbf{j}(\mathbf{x})=\frac{\hbar}{m}\mbox{Im}\left(\psi^* (\mathbf{x})\vec{\nabla}\psi(\mathbf{x}) \right)-\frac{e}{mc}\mathbf{A}|\psi|^2= \frac{\hbar}{m}\rho\nabla(S/\hbar)-\left(\frac{e\rho}{mc} \right)\mathbf{A}.
\ee
However, in our discrete setting, we must re-derive the expression $\mathbf{j}$ as a function of $n$. Note that $\mathbf{j}$ should always lie within the molecular plane. This, together with the eigenphases of $C_6$, suggests vorticity as explained in an experimental paper using electromagnetic waves \cite{stegmann_vortices_2019}. A discrete version of the continuity equation is derived below using the tight-binding Hamiltonian (\ref{eq5}) from which we can elucidate the discrete current term  and the charge density; in our case they are related with the probability flux and the probability density $|\Psi|^2$ respectively. We start from solutions of the time-dependent Schr\"odinger equation and its complex conjugate
\be
\Psi^* \left(i\hbar \frac{\partial}{\partial t}-H \right)\Psi=0, \quad \Psi\left(-i\hbar \frac{\partial}{\partial t}-H^* \right)\Psi^*=0.
\ee
Upon addition we obtain
\be\label{eq26}
i\hbar \frac{\partial}{\partial t}|\Psi|^2 = \Psi^* H \Psi - \Psi H^* \Psi^*.
\ee
In the discrete case an eigenstate of the system $|\Psi,t\>$ can be written as a linear combination of a complete basis $\{|n\>\}$ of atomic states localized at the vertices of the hexagonal ring, thus we write
\be
|\Psi,t\>=\sum_{n=1}^6 \psi_n(t)|n\>,
\ee
and eq. (\ref{eq26}) can be reexpressed using the Hamiltonian (\ref{eq5}) as follows, for each $\psi_n$:
\bea
i\hbar \frac{\partial}{\partial t}|\psi_n (t)|^2&=\psi_n^*(t)\left[E_1 \psi_n(t)+\sum_{k=1}^3(\Delta_k\psi_{n+k}(t)+\Delta^*_k\psi_{n-k}(t)) \right]\nonumber\\
&-\psi_n(t)\left[E_1\psi_n^*(t)+\sum_{k=1}^3(\Delta_k^*\psi_{n+k}^*(t)+\Delta_k\psi_{n-k}^*(t)) \right]\nonumber\\
&=\sum_{k=1}^3\left[\Delta_k\psi_n^*(t)\psi_{n+k}(t)+\Delta_k^*\psi_n^*(t)\psi_{n-k}(t)\right. \nonumber\\
&\qquad\quad \left. -\Delta_k^*\psi_n(t)\psi_{n+k}^*(t) -\Delta_k\psi_n(t)\psi_{n-k}^*(t) \right].
\eea
Assuming complex couplings $\Delta_{k}=|\Delta_k|e^{i\phi_k}$ to include the magnetic field, we have
\bea\label{eq29}
i\hbar \frac{\partial}{\partial t}|\psi_n (t)|^2=&\sum_{k=1}^3|\Delta_k|\left\{e^{i\phi_k}\left[\psi_n^*(t)\psi_{n+k}(t)-\psi_n(t)\psi_{n-k}^*(t) \right]\right. \nonumber\\
&\qquad\left. +e^{-i\phi_k}\left[\psi_n^*(t)\psi_{n-k}(t)-\psi_n(t)\psi_{n+k}^*(t) \right] \right\}.
\eea
It is important to mention at this point that our Hamiltonian is a Hermitian operator acting on a six dimensional Hilbert space and that the divergence theorem in a discrete setting must contain a sum over the ring sites. From eq. (\ref{eq29}) we may now verify the discrete version of the continuity equation: upon summation over the sites $n$, the l.h.s of (\ref{eq29}) vanishes in the stationary case \ie conservation of probability

\bea
i\hbar \sum_{n=1}^6  \frac{\partial}{\partial t} |\psi_n(t)|^2=i\hbar \frac{d}{dt}Q=0
\eea
where $Q$ is the total probability. On the other hand the r.h.s of (\ref{eq29}) vanishes by hermiticity of the Hamiltonian; this is proved below
\bea
i\hbar \frac{\partial}{\partial t} \sum_{n=1}^6 |\psi_n(t)|^2=\sum_{k=1}^3|\Delta_k|\left\{ e^{i\phi_k}\sum_{n=1}^6\left[\psi_n^*(t)\psi_{n+k}(t)-\psi_n(t)\psi_{n-k}^*(t) \right] \right. \nonumber\\
\left.+e^{-i\phi_k}\sum_{n=1}^6\left[\psi_n^*(t)\psi_{n-k}(t)-\psi_n(t)\psi_{n+k}^*(t) \right] \right\} \nonumber
\eea
\bea
=\sum_{k=1}^3\sum_{n=1}^6|\Delta_k|\left\{e^{i\phi_k}\left[\,|\psi_n\psi_{n+k}|e^{i(\chi_{n+k}-\chi_n)}-|\psi_n\psi_{n-k}|e^{i(\chi_n-\chi_{n-k})}\right] \right. \nonumber\\
\qquad\qquad\quad\left. +e^{-i\phi_k}\left[\,|\psi_n\psi_{n-k}|e^{i(\chi_{n-k}-\chi_n)}-|\psi_n\psi_{n+k}|e^{i(\chi_n-\chi_{n+k})} \right] \right\}\nonumber\\
=\sum_{n,m=1}^6 \left(\psi_n H^*_{mn}\psi_m^*-\psi_n H_{nm}\psi_m^* \right)=0
\eea
This indicates that the r.h.s. of (\ref{eq29}) should be expressed as a current term of a discrete continuity equation
\be\label{eq30}
i\hbar \sum_{k=1}^3(j_{n+k}^{(k)}-j_n^{(k)})\alpha_k
\ee
which always vanishes when
\be
\sum_{n=1}^6j_{n+k}=\sum_{n=1}^6j_n, \qquad j_n=\sum_{k=1}^3 j_n^{(k)},
\ee
for $k=1,2,3$ because of periodicity (mod $6$). This is in full parallel with the divergence theorem. Some algebra must be done to find the exact expression for $j_n^{(k)}$. Re-expressing the r.h.s. of (\ref{eq29})
\bea
&\sum_{k=1}^3|\Delta_k|\left\{e^{i\phi_k}\left(\psi_n^*\psi_{n+k}-\psi_n\psi_{n-k}^* \right)+e^{-i\phi_k}\left(\psi_n^*\psi_{n-k}-\psi_n\psi_{n+k}^* \right) \right\}\nonumber\\
&=\sum_{k=1}^3|\Delta_k|\left\{\left(e^{i\phi_k}\psi_n^*\psi_{n+k}-e^{-i\phi_k}\psi_n\psi_{n+k}^* \right)\right. \nonumber \\
&\left. -\left(e^{i\phi_k}\psi_{n-k}^*\psi_n-e^{-i\phi_k}\psi_{n-k}\psi_n^* \right) \right\}
\eea
we note that it coincides with (\ref{eq30}) if
\bea
|\Delta_k|=\alpha_k\quad \mbox{\small and}\quad j_n^{(k)}&=\frac{1}{i\hbar}\left(e^{i\phi_k}\psi_{n-k}^*\psi_n-e^{-i\phi_k}\psi_{n-k}\psi_n^* \right)\nonumber\\
&=\frac{2}{\hbar}\mbox{\small Im} \left(e^{i\phi_k}\psi_{n-k}^*\psi_n \right).
\eea
For eigenfunctions, we trivially recover
\be
\frac{\partial}{\partial t}\left(|\psi_n(t)|^2 \right)=0 \quad \mbox{\small and} \sum_{k=1}^3 (j_{n+k}^{(k)}-j_n^{(k)})\alpha_k=0.
\ee
Now we can write the stationary wave function as $\psi_n(t)=\varphi_n e^{-iE_q t /\hbar}$ in order to obtain an expression for the current which is independent of $t$
\be
j_n=\frac{2}{\hbar}\mbox{\small Im} \left(\sum_{k=1}^3 e^{i\phi_k}\varphi_{n-k}^*\varphi_n \right).
\ee
This represents a stationary current related to  $k$ neighbours. The currents are plotted in fig. \ref{fig4.1} for the particular case of nearest neighbours ($k=1$) and $\varphi_n\in \mathbb{R} $. The length of the arrow lines represent the intensity of the current. The maximum value is reached when the phase is $\phi=\pi/2$.

\bfig
\bcen
\includegraphics[width=15cm]{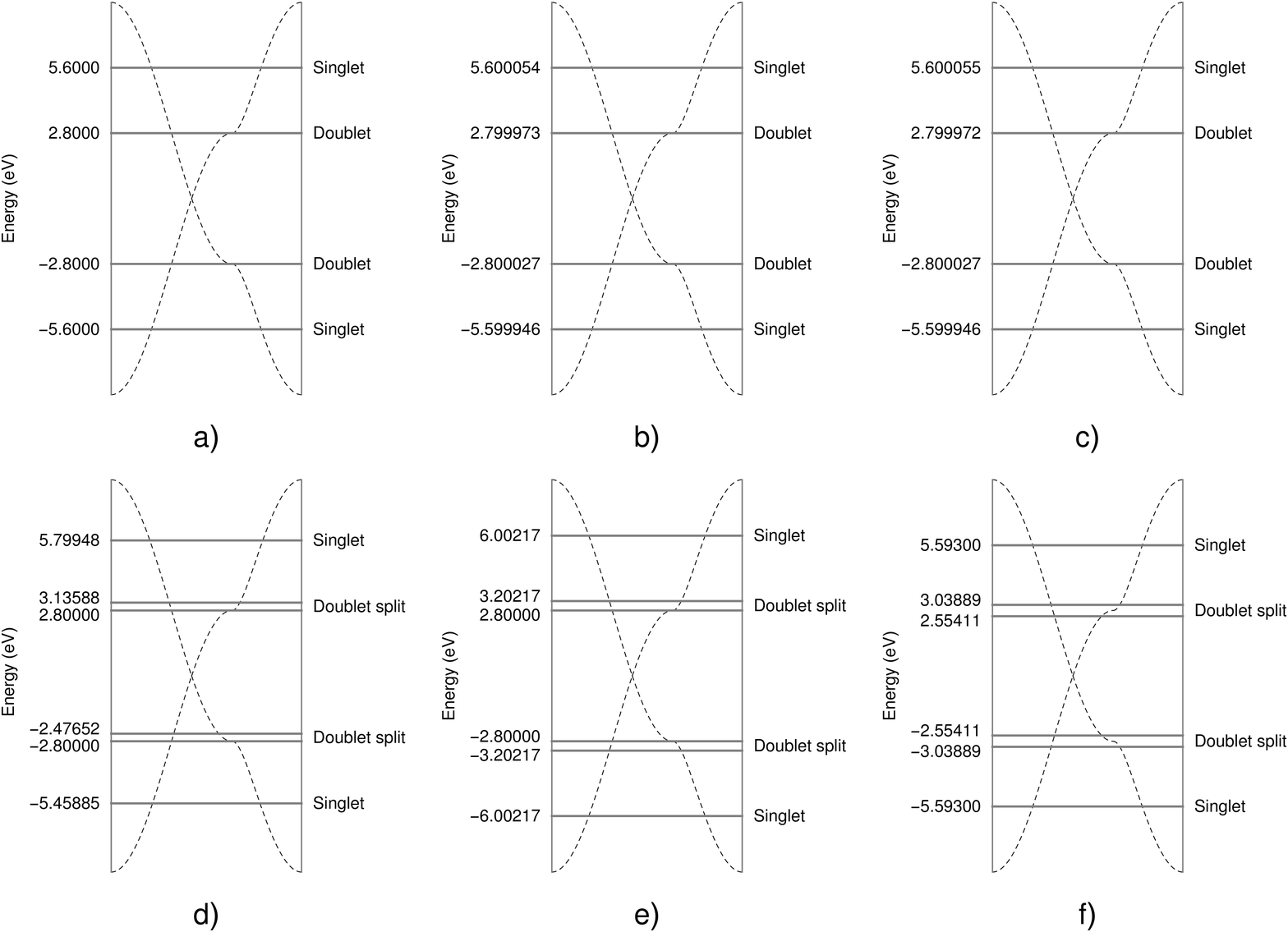}
\caption{\label{fig4} Energy levels for benzene. The distance between sites is taken as $1.39$ \AA\,  and the on site energy is shifted to $0$. a) Nearest-neighbour couplings. The energy structure (from top to bottom) is singlet, doublet, doublet, singlet. The energy levels posses a reflection symmetry. b) Second neighbour couplings break reflection symmetry.  c) Third nearest-neighbour couplings. The corrections to energy levels due to second and third neighbours are negligible.  d) Nearest neighbour couplings with a site defect of $1$ eV. e) Nearest neighbour couplings with coupling defects $(1,2)$, $(1,6)$, $(6,1)$ and $(2,1)$ by $0.3$ eV. f) Nearest neighbour couplings with external magnetic field perpendicular to the ring plane of $B=8.5855\times 10^{-3}$G. }
\ecen
\efig

\subsection{Numerical results}

\bfig
\bcen
\includegraphics[width=15cm]{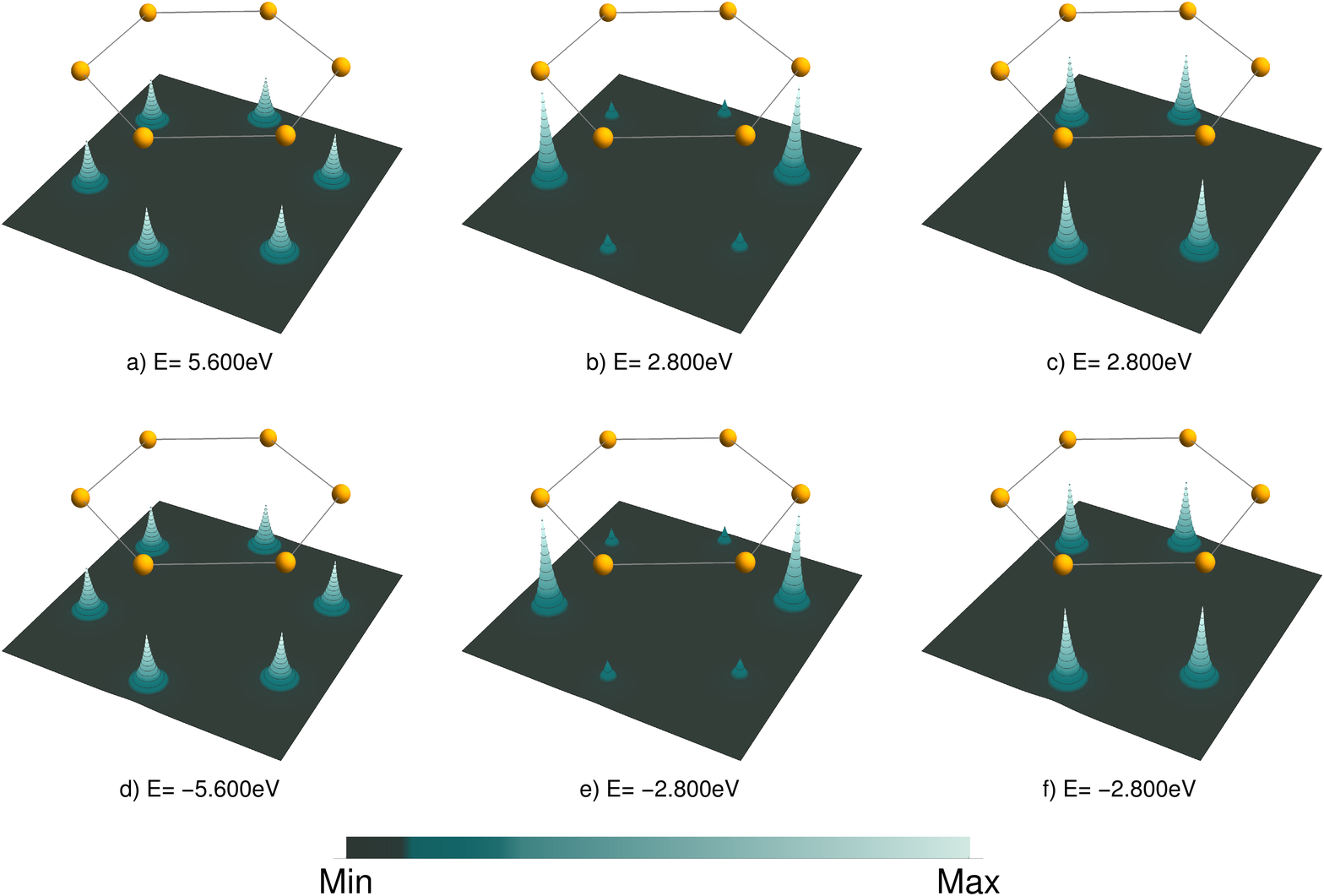}
\caption{\label{fig5} Eigenfunctions for a nearest neighbour tight-binding model of benzene. Panels a) and d) represent the singlet states and display the same probability of finding the electron in any of the six sites of the ring. The benzene ring is presented as a visual reference of the spatial dependence of the eigenfuntions. }
\ecen
\efig

\bfig
\bcen
\includegraphics[width=15cm]{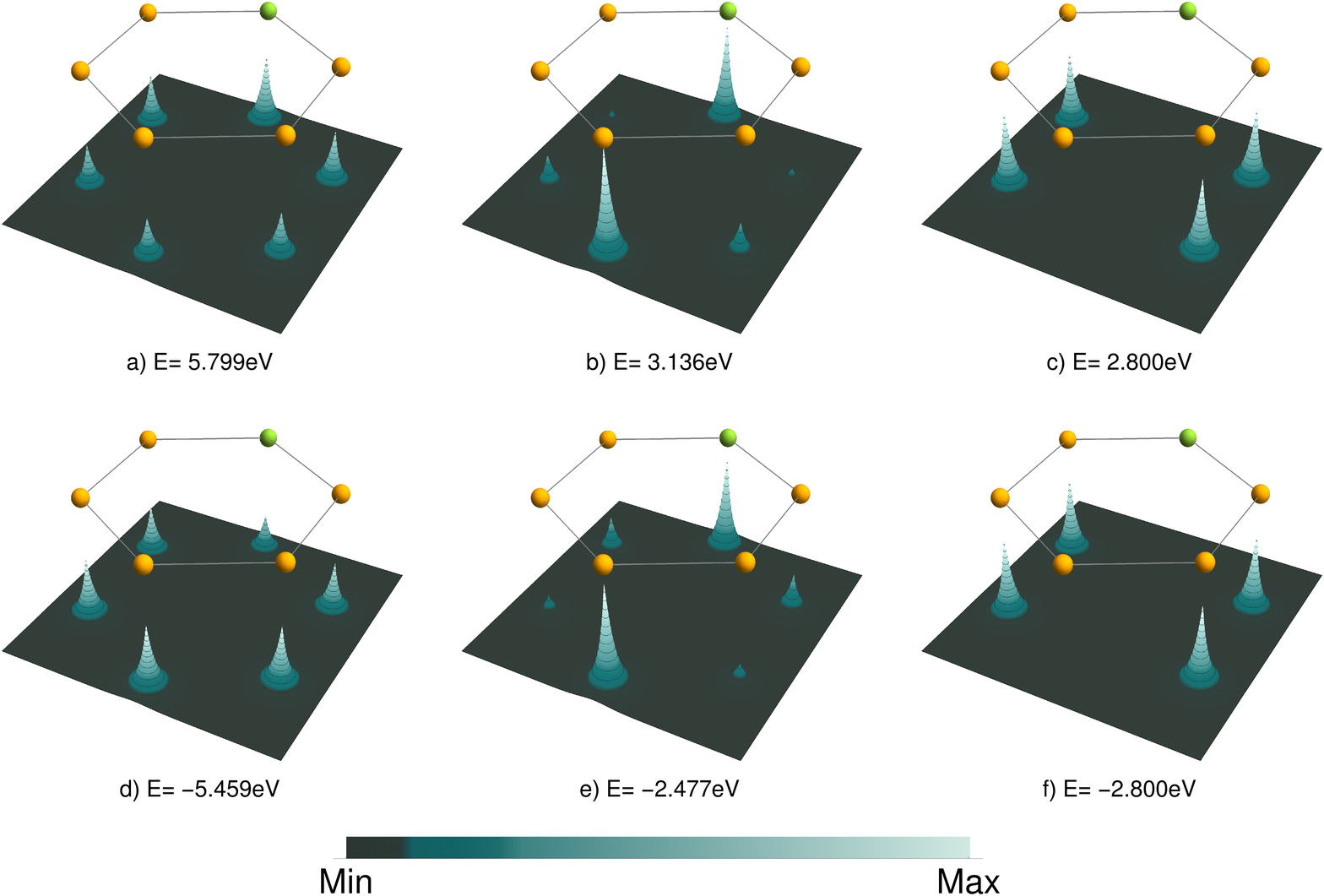}
\caption{\label{fig6} Eigenfuntions of a benzene ring with $C_6$ symmetry broken by one site defect of $1$eV. The spectrum is depicted in figure \ref{fig4} panel d). Panels a) and d) show singlet eigenfunctions, the probability is modified in the region around the defect site. In panels b) and e) depict the modification to the probability is presented in two sites of the ring while in panels c) and f) no visually change is present.   }
\ecen
\efig

This section presents some calculations for benzene. The necessary data, such as valence energy, atomic distances and bond angles are well-known and reported in several handbooks and articles \cite{engelberts_electronic_2005,Harshbarger_1969,Watanabe_2004,Geim_2007,Castro_Neto_2009,Zhao_2010}.

Since we are only interested in the distance dependence of the couplings, we focus our interest in the radial part of the atomic wave function. Regarding the molecular structure and the well-known bonding structure in benzene as a reasonable estimate, we restrict ourselves to the $p_z$ orbital.  The radial dependence of $p_z$  has the following form:
\be
R(r)=Are^{-r/\lambda}, \quad \lambda=\frac{2a_0}{Z},\quad A=\left(\frac{Z}{2a_0} \right)^{3/2} \frac{Z}{\sqrt{3}a_0},
\ee
where $a_0$ stands for the Bohr radius and $Z$ for the nuclear charge \footnote{Other corrections to the Bohr radius may arise from the atomic shielding, which modifies the evanescence length of the wave function \cite{Slater_1930,Snyder_1971}. Nevertheless, for the purpose of the present work, this does not introduce major corrections and can be dismissed.}. The couplings are given by the overlap integral in eq. (\ref{eq2}). A reasonable approximation of the wave function in regions outside the atom is $R(r)\sim e^{-r\lambda}$, therefore the couplings take the form
\be
\Delta \sim \Delta_0 e^{-d/\tilde{\lambda}},\quad \tilde{\lambda}=\frac{\lambda_1 \lambda_2}{\lambda_1 + \lambda_2}
\ee
where $d$ is the distance between the two atoms; $\lambda_1$ and $\lambda_2$  are the evanescence lengths for atoms $1$ and $2$ correspondingly and $\Delta_0$ is a constant. For benzene $\lambda_C=a_0/3$  while for borazine $\lambda_N=2a_0/7$ and $\lambda_B=2a_0/5$. In order to obtain numerical values of $\Delta_0$ and site energies, we may take the information from parameters of graphene in the case of benzene and boron nitride for borazine \cite{Geim_2007,Castro_Neto_2009,Zhao_2010}.

Figure \ref{fig4} shows the resulting energy levels obtained from our model. Panel a) illustrates a first neighbour coupling spectrum where the up-down symmetry is present; on the other hand, panel b) shows the spectrum of benzene including second neighbours, where the up-down symmetry is visibly broken. Panel c) corresponds to the full spectrum with all pairwise couplings. The corrections to the level structure generated by third neighbour couplings are not quantitatively relevant and we may conclude that a nearest neighbour model suffices for the description of the benzene ring.  Panels d), e) and f) illustrate the level spiting of doublets via three different ways:
\begin{itemize}
\item[ (i) ]  Introducing a defect in one site. See panel d) of figure \ref{fig4}.
\item[ (ii) ]  Introducing coupling defects in $\Delta_{(1,2)}$, $\Delta_{(1,6)}$, $\Delta_{(2,1)}$ and $\Delta_{(6,1)}$. Figure \ref{fig4} panel e).
\item[ (iii) ]  Introducing a magnetic field through the benzene ring. See panel f) of figure \ref{fig4}.
\end{itemize}
The eigenfunctions are plotted in figure \ref{fig5} for a nearest neighbour model and for one site defect Hamiltonian in figure \ref{fig6}. A ring is shown as a visual reference. In artificial realizations, the evanescence length $\lambda$ can be obtained from previous knowledge of localized states at discs, corners or junctions. An interesting case arises when $\lambda$ is comparable to the bonding length where a triple degeneracy point was found in a second neighbour coupling model  when $\Delta_2/\Delta_1=1/3$. In the third neighbour coupling model, the triple degeneracy point emerges when $1-3(\Delta_2/\Delta_1)+4(\Delta_3/\Delta_1)=0$. This doublet-singlet degeneracy has been studied previously by means of geometric deformations \cite{Sadurni_2019}.

\bfig
\bcen
\includegraphics[width=12cm]{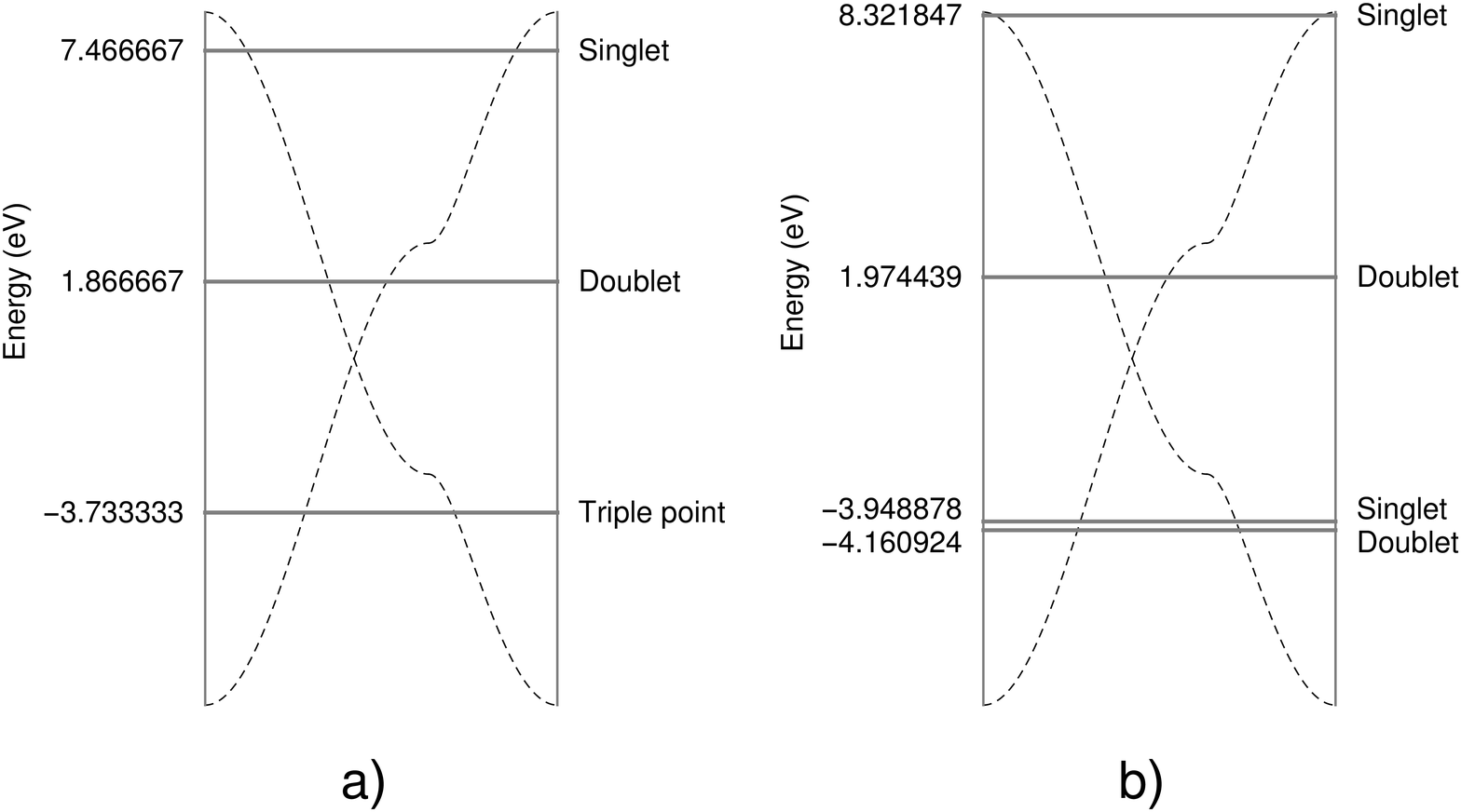}
\caption{ Second neighbour couplings spectra. a) Triple point between inferior singlet and doublets corresponding to $\Delta_2/\Delta_1=1/3$ and  $\lambda_c=0.926$\AA. b) Inferior doublet-singlet inversion. The level inversion occurs when $\Delta_2/\Delta_1>1/3$ or $\lambda>\lambda_c$ in this case $\lambda=0.956$\AA. }
\ecen
\efig

\bfig
\bcen
\includegraphics[width=15cm]{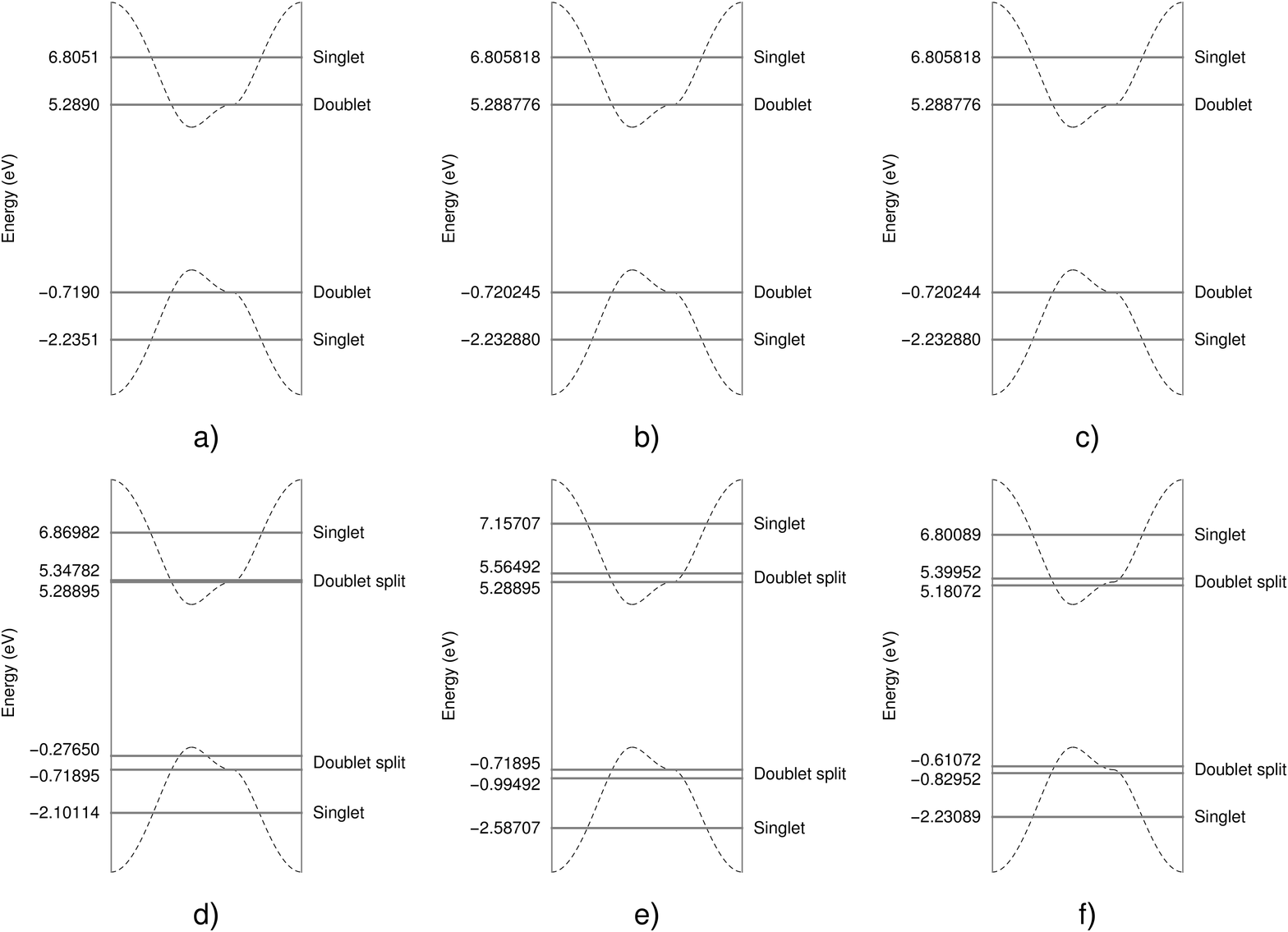}
\caption{\label{fig4} Energy levels for borazine. The distance between sites is taken as $1.43$ \AA\,  and the on site energies as $\epsilon_1=0$eV and $\epsilon_2=4.57$eV. a) Nearest-neighbour couplings. The energy structure (from top to bottom) is singlet, doublet, doublet and singlet thus reflection symmetry emerges as in benzene. b) Second neighbour couplings. c) Third nearest neighbour couplings. Panels d), e) and f) show doublets lift.  d) Nearest neighbour couplings with a site defect by an increment of $0.7$eV in one boron site.  e) Nearest neighbour couplings with coupling defects by an increment of $0.3$eV in $\Delta_{(1,2)}$, $\Delta_{(1,6)}$, $\Delta_{(6,1)}$ and $\Delta_{(2,1)}$. f) Nearest neighbour couplings with an external magnetic field $B=8.5855\times10^{-3}$G. }
\ecen
\efig

\bfig
\bcen
\includegraphics[width=15cm]{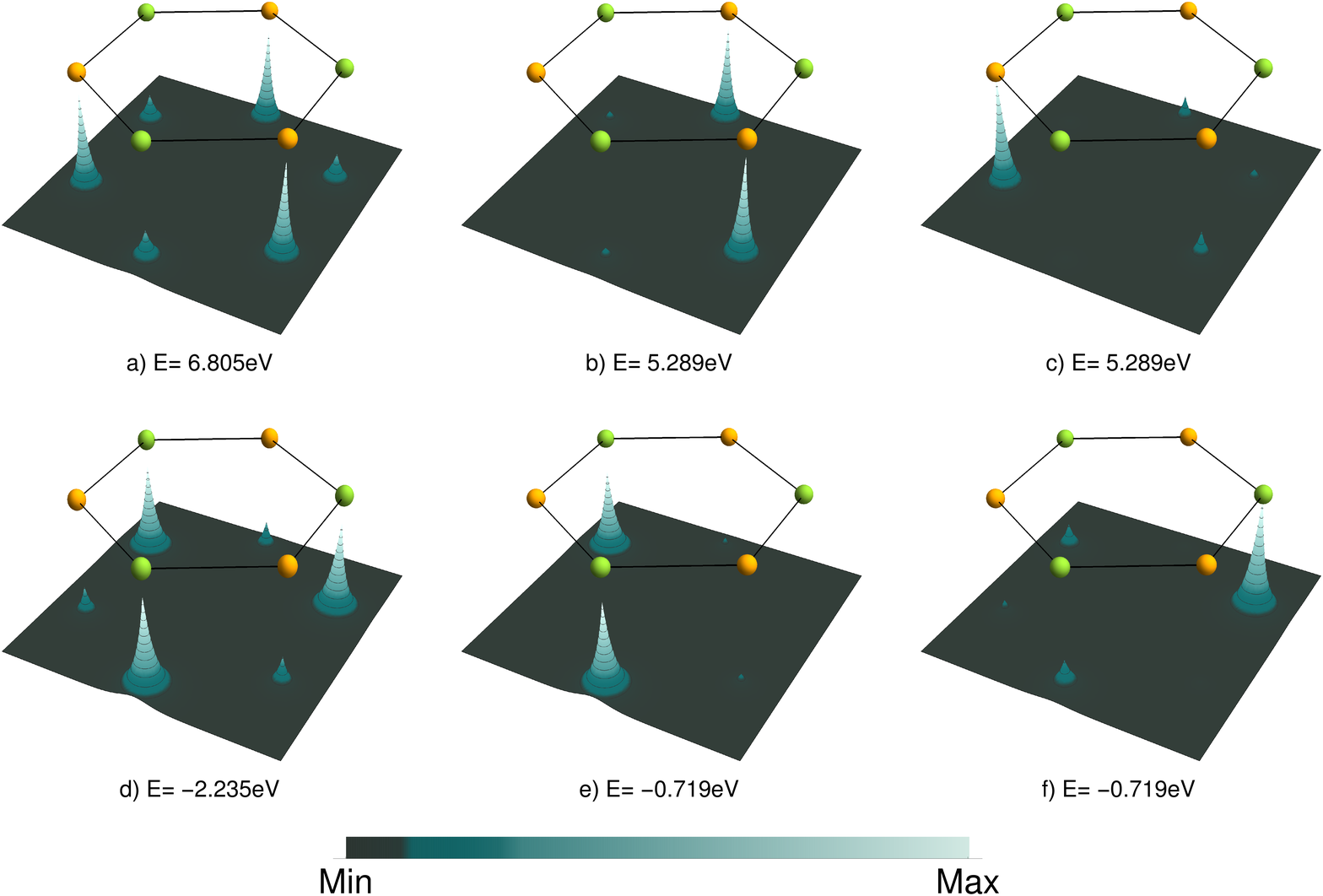}
\caption{\label{fig5} Eigenfunctions for a nearest neighbour tight-binding model of borazine. Yellow spheres represent nitrogen atoms and green boron atoms. Panels a) and d) show the singlet states. Panels b), c) and e), f) show the wave functions for degenerated  doublets. }
\ecen
\efig

\bfig
\bcen
\includegraphics[width=15cm]{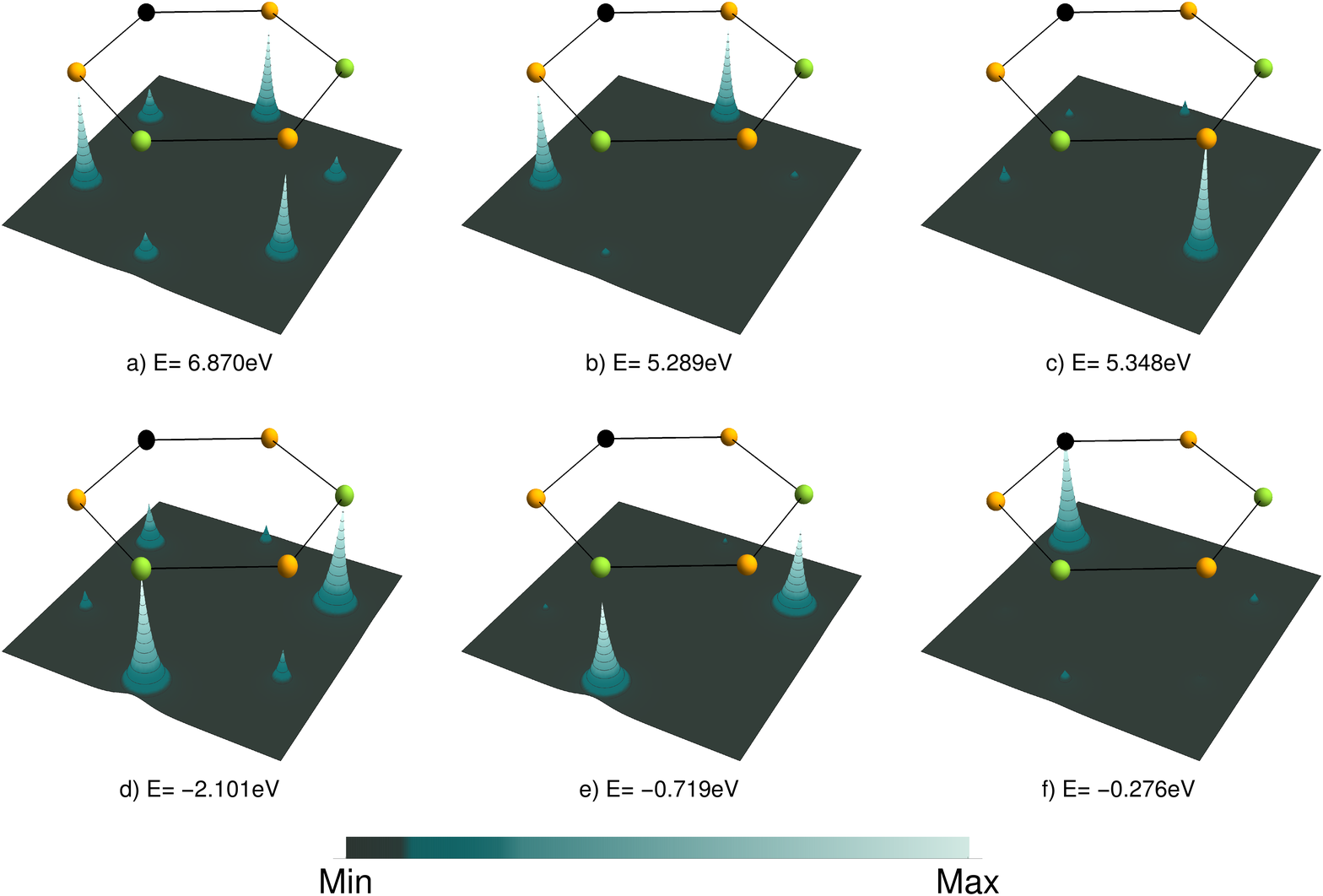}
\caption{\label{fig6} Eigenfuntions of  borazine ring with $C_3$ symmetry broken by one site defect increment of $0.7$ eV. Yellow spheres represent nitrogen atoms, green boron and the black spheres is for the on-site defect. Panels a) and d) stand for the singlet levels. Panels b), c) and e), f) illustrate the wave functions of the lifted doublets.   }
\ecen
\efig

\section{Borazine \label{sec:2}}
\subsection{Multiple neighbours in a $C_3$ scheme}
In this section we introduce a system with only a (partial) triangular symmetry, in contrast with benzene, as it is the appropriate model for a borazine molecule. The group $C_6$ can be decomposed into $C_3$ and $C_2$: $C_6 =C_3 \times C_2 = \{\mathbf{1},\tau,\tau^2 \}\times \{\mathbf{1},P \}$ such that $P^2 = \mathbf{1}$ and $\tau^3=\mathbf{1}$. Evidently $T^2=\tau$, as it represents a rotation of $2\pi/3$ radians around the centre of a hexagon, and $P$ swaps boron and nitrogen sites and vice versa. Only $N$ (nitrogen) centres allow shuffling; similarly, only B (boron) centres allow shuffling among themselves. The on-site energies for valence orbitals are different: $E_B=\epsilon_1$, $E_N=\epsilon_2$. As to the couplings, the $C_3$ symmetry is preserved if $\Delta_1$ represents N-B couplings (nearest neighbours), $\Delta_2$ represents either B-B couplings $(\Delta_2^+)$ or N-N couplings $(\Delta_2^-)$ and $\Delta_3$ yields the N-B couplings between opposite vertices, which must be identical in all three possibilities according to the geometry. The resulting Hamiltonian is reorganized as
\be
H\doteq
\left(
\begin{array}{ccc|ccc}
\epsilon_1&\Delta_2^+&\Delta_2^+&\Delta_1&\Delta_3&\Delta_1\\
\Delta_2^+&\epsilon_1&\Delta_2^+&\Delta_1&\Delta_1&\Delta_3\\
\Delta_2^+&\Delta_2^+&\epsilon_1&\Delta_3&\Delta_1&\Delta_1\\
\hline
\Delta_1&\Delta_1&\Delta_3&\epsilon_2&\Delta_2^-&\Delta_2^-\\
\Delta_3&\Delta_1&\Delta_1&\Delta_2^-&\epsilon_2&\Delta_2^-\\
\Delta_1&\Delta_3&\Delta_1&\Delta_2^-&\Delta_2^-&\epsilon_2\\
\end{array}
\right).
\ee
The operation $P|N\>=|B\>$, $P|B\>=|N\>$ in its matrix form is represented by
\bea
P\doteq
\left(
\begin{array}{cc}
0&\mathbf{1}_{3\times3}\\
\mathbf{1}_{3\times3}&0\\
\end{array}
\right)
= \mathbf{1}_{3\times3}\otimes\sigma_1,
\eea
where $\sigma_i$ stand for the Pauli matrices. Now the matrix representation of the translation operator $\tau$ in $C_3$ is

\bea
 \tau|N_k\>=|N_{(k+1)\,\mbox{\scriptsize mod}\, 3}\>, \quad \tau|B_k\>=|B_{(k+1)\, \mbox{\scriptsize mod}\, 3}\>,\nonumber\\
\tau\doteq
\left(
\begin{array}{ccc|ccc}
0&0&1&0&0&0\\
1&0&0&0&0&0\\
0&1&0&0&0&0\\\hline
0&0&0&0&0&1\\
0&0&0&1&0&0\\
0&0&0&0&1&0\\
\end{array}\right)
=
\left(
\begin{array}{ccc}
0&0&1\\
1&0&0\\
0&1&0\\
\end{array}
\right)
\otimes
\mathbf{1}_{2\times 2}.
\eea
Let us introduce the notation $|n,m\>$ for the localised states, where $n$ represents the number of subtriangle (1 for boron and 2 for nitrogen) and the index $m$ represents the internal position in each subtriangle. In this notation, $P$ and $\tau$ operate as follows
\bea
P |n,m\>=|(n+1)\, \mbox{\small mod}\, 2,m\>,\nonumber\\
\tau|n,m\>=|n,(m+1) \, \mbox{\small mod}\, 3\>,\nonumber\\
\left[ \tau,P \right]=0.
\eea
The Hamiltonian must be expressed as a sum of operators, therefore we need to define other matrices:
\be
Q_\pm=\mathbf{1}_{3\times 3}\otimes \sigma_\pm \quad \mbox{\small and} \quad \Pi_\pm=\mathbf{1}_{3\times3}\otimes(1/2)(\mathbf{1}_{2\times2}\pm\sigma_3),
\ee
where $\sigma_\pm=(1/2)(\sigma_1\pm i\sigma_2)$. $H$ can be expressed now as function of the operators $\Pi_{\pm}$, $\tau$ and $Q_{\pm}$:
\bea\label{eq41}
H=&\Pi_+(\epsilon_1+\tau\Delta_2^+ +\tau^\dagger \Delta_2^+)+\Pi_-(\epsilon_2+\tau\Delta_2^- + \tau^\dagger\Delta_2^-)\nonumber\\
&+Q_+(\Delta_1+\Delta_1\tau+\Delta_3\tau^\dagger)+Q_-(\Delta_1+\Delta_1\tau^\dagger+\Delta_3\tau),
\eea
with the following commutation and anticommutation relations
\bea
Q_+^2 = Q_-^2 = 0, \quad [Q_\pm, \, \tau^n]=0, \quad [\Pi_\pm, \, \tau]=0 \nonumber\\
\left[Q_+, \, Q_- \right]=\mathbf{1}_{3\times3}\otimes[\sigma_+,\sigma_-]=\mathbf{1}_{3\times3}\otimes\sigma_3=\Pi_+ -\Pi_- =\Sigma_3\nonumber\\
\left\{Q_+,Q_- \right\}=\mathbf{1}_{3\times3}\otimes\{\sigma_+,\sigma_- \}=\Pi_+ + \Pi_-=\mathbf{1}_{6\times6}\nonumber\\
\left[Q_\pm, \, P\right]=\mathbf{1}_{3\times3}\otimes[\sigma_\pm,\sigma_1]=\pm i \mathbf{1}_{3\times3}\otimes[\sigma_2,\sigma_1]=\mp2 i^2 \mathbf{1}_{3\times3}\otimes \sigma_3=\pm2\Sigma_3\nonumber\\
\left[Q_\pm, \, \Pi_\pm \right]=\mathbf{1}_{3\times3}\otimes[\sigma_\pm, \, \pm \frac{1}{2}\sigma_3]=\pm\frac{1}{2}\mathbf{1}_{3\times3}\otimes(\mp 2 \sigma_\pm)=\pm(\mp Q_\pm)\nonumber\\
\left\{ Q_\pm, \, \Pi_\pm \right\}=\mathbf{1}_{3\times3}\otimes \left\{\sigma_\pm, (\mathbf{1}_{2\times2}\pm \sigma_3)/2 \right\}=1_{3\times3}\otimes\sigma_\pm=Q_\pm
\eea
This problem can be happily solved in a simple way by using the eigenphases of $C_3$, which turn out to be
\bea
|l\>=\frac{1}{\sqrt{2}}\frac{1}{\sqrt{3}}\sum_{n=1}^2\sum_{m=1}^3 e^{-i\frac{2\pi lm}{3}}|n,m\>,\quad l=1,2,3 \nonumber\\
|l,n\>=\frac{1}{\sqrt{3}}\sum_{m=1}^3e^{-i\frac{2\pi lm}{3}}|n,m\>,\quad \<l',n'|l,n\>=\delta_{ll'}\delta_{nn'}
\eea
The elements of the basis $\{|l\>\}$ are eigenstates of the translation operator $\tau$
\bea
\tau|l\>=e^{i\frac{2\pi l}{3}}|l\>, \, l=1,2,3.
\eea
The Hamiltonian operates on $|l\>$ as below
\bea
H|l\>=&\left\{\Pi_+\left[\epsilon_1+2\Delta_2^+ \cos\left(\frac{2\pi l}{3}\right)\right]+\Pi_- \left[\epsilon_2+2\Delta_2^- \cos\left(\frac{2\pi l}{3}\right)\right]\right.\nonumber\\
&+Q_+\left[\Delta_1+\Delta_1 e^{i\frac{2 \pi l}{3}}+\Delta_3 e^{-i\frac{2\pi l}{3}}\right]\nonumber\\
&\left.+Q_-\left[\Delta_1 + \Delta_1 e^{-i\frac{2\pi l}{3}}+\Delta_3 e^{i\frac{2\pi l}{3}}\right] \right\}|l\>.
\eea
Now, the operators $Q_{\pm}$ and $\Pi_{\pm}$ acting on $|l\>$ result in the following

\bea
\Pi_+|l\>=\frac{1}{\sqrt{3}}\frac{1}{\sqrt{2}}\sum_{m=1}^3 e^{i\frac{2\pi lm}{3}}|1,m\>=\sqrt{2}|l,1\>\nonumber\\
\Pi_-|l\>=\frac{1}{\sqrt{3}}\frac{1}{\sqrt{2}}\sum_{m=1}^3 e^{i\frac{2\pi lm}{3}}|2,m\>=\sqrt{2}|l,2\>\nonumber\\
Q_+|l\>=\sqrt{2}|l,1\>\nonumber\\
Q_-|l\>=\sqrt{2}|l,2\>\nonumber\\
\tau|l,n\>=e^{i\frac{2\pi l}{3}}|l,n\>.
\eea
Finally the Hamiltonian operates on $\{|l,n\>\}$ as

\bea
H|l,1\>=\left[\epsilon_1+2\Delta_2^+\cos\left(\frac{2\pi l}{3} \right)\right]|l,1\>+\left[\Delta_1+\Delta_1  e^{-i\frac{2\pi l}{3}}+\Delta_3 e^{i\frac{2\pi l}{3}}\right]|l,2\>\nonumber\\
H|l,2\>=\left[\epsilon_2+2\Delta_2^-\cos\left(\frac{2\pi l}{3} \right)\right]|l,2\>+\left[\Delta_1+\Delta_1  e^{i\frac{2\pi l}{3}}+\Delta_3 e^{-i\frac{2\pi l}{3}}\right]|l,1\>,\nonumber\\
\eea
leading to the following matrix representation:
\be
H^{(l)}\doteq
\left(
\begin{array}{cc}
\epsilon_1+2\Delta_2^+\cos\left(\frac{2\pi l}{3}\right)&\Delta_1+\Delta_1 e^{-i\frac{2\pi l}{3}}+\Delta_3 e^{i\frac{2\pi l}{3}}\\
\Delta_1+\Delta_1 e^{i\frac{2\pi l}{3}}+\Delta_3 e^{-i\frac{2\pi l}{3}}&\epsilon_2 + 2\Delta_2^- \cos\left(\frac{2\pi l}{3} \right)\\
\end{array}
\right)
\ee
where $l=1,2,3$ and which is solvable by radicals. The wave functions are explicitly given by 

\be
N^{(l)}\left[|l,1\>-A^{(l)} |l,2 \> \right], \quad N^{(l)}\left[|l,1\> +( A^{(l)}) ^* |l,2\> \right],
\ee
where, in terms of matrix elements $H^{(l)}_{ij}$ above, we write 
\be
N^{(l)}=\frac{1}{\sqrt{1+|A^{(l)}|^2}},\quad A^{(l)}=\frac{\frac{H_{22}^{(l)}-H_{11}^{(l)}}{2}+\sqrt{\left( \frac{H_{22}^{(l)}-H_{11}^{(l)}}{2}  \right)^2 +|H_{12}^{(l)}|^2}}{(H_{12}^{(l)})^*}.
\ee

These expressions can be regarded as analytical solutions which, to our knowledge, have not been provided for borazine elsewhere. 

\section{Supersymmetry \label{sec:3}}

The two copies of a triangle spectrum that appear reflected in benzene and borazine can be explained by finding the dynamical supersymmetry (SUSY) of the ring. For the Dirac oscillator, the supersymmetry is well-known, as the reader may verify in \cite{sadurni_diracmoshinsky_2011,castanos_soluble_1991}. In our present case, a similar treatment can be developed in order to define a superalgebra. Let us rewrite the Hamiltonian (\ref{eq41}) as follows:
\bea
H=\hat{\Pi}_+ \left(\epsilon_1 + \tau \Delta_2^+ + \tau^\dagger \Delta_2^+ \right) + \hat{\Pi}_- \left(\epsilon_2 + \tau \Delta_2^- + \tau^\dagger \Delta_2^- \right)+ \hat{C}_1\nonumber\\
=\hat{\epsilon}_0+\hat{\mu}\Sigma_3 +\hat{C}_1,
\eea
where the operators $\hat{C}_1$, $\hat{\epsilon}_0$ and $\hat{\mu}$ have been defined as
\bea
\hat{C}_1=Q_+ \left(\Delta_1 + \Delta_1 \tau + \Delta_3 \tau^\dagger \right)+ Q_- \left(\Delta_1 + \Delta_1 \tau^\dagger + \Delta_3 \tau \right)\nonumber\\
\hat{\epsilon}_0=\frac{1}{2}\left[(\epsilon_1 + \epsilon_2)\mathbf{1}_{6\times6} + \tau\Delta_2^+ + \tau^\dagger\Delta_2^+ +  \tau\Delta_2^- + \tau^\dagger\Delta_2^- \right]\nonumber\\
\hat{\mu}=\frac{1}{2}\left[(\epsilon_1 - \epsilon_2)\mathbf{1}_{6\times6} + \tau\Delta_2^+ + \tau^\dagger\Delta_2^+ - \tau\Delta_2^- - \tau^\dagger\Delta_2^-\right].
\eea
Following the references \cite{sadurni_diracmoshinsky_2011} we define the operator $\hat{C}_2$ as:
\be
\hat{C}_2=-i Q_+ \hat{a}+iQ_- \hat{a}^\dagger,
\ee
and the ``bosonic'' definitions of $a$ and $a^\dagger$ are shown below:

\bea
\hat{a}=\Delta_1 + \Delta_1\tau + \Delta_3\tau^\dagger, \quad
\hat{a}^\dagger=\Delta_1 + \Delta_1\tau^\dagger + \Delta_3\tau,\quad [\hat{a},\hat{a}^\dagger]=0 &.
\eea
We verify the (anti)commutation relations between $\hat{C}_1$ and $\hat{C}_2$
\bea
\{\hat{C}_1, \, \hat{C}_2 \}=0,\nonumber\\
\left[\hat{C}_i, \, \{ \hat{C}_j,\hat{C}_k \} \right]=0,\quad i,j,k=1,2,3 \nonumber\\
\{\hat{C}_1,\,\hat{C}_1 \}=\{\hat{C}_2,\,\hat{C}_2 \}=2\left(\hat{a}\hat{a}^\dagger \Pi_+ + \hat{a}^\dagger \hat{a} \Pi_- \right),\nonumber\\
\{\hat{\mu}\Sigma_3,\, \hat{C}_1 \}=\Sigma_3 Q_+ \left[\hat{\mu},\hat{a} \right]+\Sigma_3 Q_- \left[\hat{\mu},\hat{a}^\dagger \right].
\eea
Now, if we have that $\hat{a}$ and $\hat{\mu}$ commute, a central charge $\left(H-\hat{\epsilon}_0 \right)^2-\hat{\mu}^2$ can be defined and the following SUSY is confirmed:
\bea
\left(H-\hat{\epsilon}_0\right)^2=\left(\hat{\mu}\Sigma_3 + \hat{C}_1 \right)^2=\hat{\mu}^2 + \hat{C}_1^2\nonumber\\
\left(H-\hat{\epsilon}_0 \right)^2-\hat{\mu}^2=aa^\dagger \Pi_+ + a^\dagger a \Pi_-\nonumber\\
\{\hat{C}_1,\hat{C}_1 \}=\{\hat{C}_2,\hat{C}_2 \}=2\left[\left(H-\hat{\epsilon}_0 \right)^2 - \hat{\mu}^2 \right]\nonumber\\
\{\hat{C}_i,\hat{C}_j \}=2\delta_{ij}\left[\left(H-\hat{\epsilon}_0 \right)^2 - \hat{\mu}^2 \right].
\label{susy}
\eea

Since this $S(2)$ superalgebra is dynamical, we do not expect the charges $C_{1,2}$ to relate bosonic and fermionic sectors directly. Instead, they connect energy states below the central level (here taken as $\epsilon=0$) with those above. In a general sense, the supercharges relate {\it dressed states,\ }made of dimers and trimers. When we look for the operators that relate {\it undressed\ }bosonic and fermionic sectors, we encounter $Q_\pm$ as responsible for such operation. In particular, $a, a^\dagger$ are an abelian realization of ladder operators that are admissible in our finite $3\times 3$ representation, while $\Sigma_i$ constitute the {\it undressed\ }fermionic sector, as obviously shown by $\{\Sigma_i , \Sigma_j \}=2\delta_{i,j}\v 1_{4\times 4}$. Some interesting consequences can be studied from the model above:
\begin{itemize}
\item If $[\hat{\mu}, \hat{a}]\neq 0$ (and as a consequence $[\hat{\mu}, \hat{a}^\dagger]\neq 0$) then the relation $\{\hat{C}_i, \hat{C}_j \}=2\delta_{ij}\left[\left(H - \hat{\epsilon}_0 \right)^2-\hat{\mu}^2 \right]$ breaks and the term $\left(H-\hat{\epsilon}_0 \right)^2-\hat{\mu}^2$ can not be used as central charge any more, breaking thereby the SUSY.
\item If $[\hat{a},\hat{a}^\dagger]\neq 0$ (Heisenberg and other Cartan algebras, for example) the dynamical SUSY is preserved.
\item For each subtriangle, if the symmetry group $C_3$ holds then $\hat{\mu}\propto \mathbf{1}$ and $[\hat{\mu},a]=0$, but if we have intertriangular couplings that break $C_3$ globally, then $[a, a^\dagger]\neq 0$ in general. The supersymmetry is preserved even if we break the global $C_3$, but not the symmetry of each subtriangle.
\end{itemize}
Due to $\left(H-\hat{\epsilon}_0 \right)^2 -\hat{\mu}^2$, the dynamical supersymmetry yields reflection symmetry between energy pairs with variable geometric centres according to $l=1,2,3$. Such centres add to $3(\epsilon_1 +\epsilon_2 )/2$ always. Since $C_3$ is not broken  in benzene or borazine, we may conclude that these compounds enjoy SUSY in their spectrum.

It is interesting to note that a generalization of the algebra (\ref{susy}) to deformed cases might be helpful to deal with asymmetrical configurations of the molecule, say, if external strain or vibrations are applied to it. In this scenario, the $C_3$ sector is no longer valid. In addition, the typical bosonic promotion  $\left[a, a^\dagger \right]=1$ does not work in finite dimension because the Heisenberg algebra is non-compact. The possibilities that can be considered here are the Cartan forms of compact algebras such as $J_{\pm}$ for su(2) for $j=1$ or hypercharge operators $U_{\pm}, V_{\pm}$ for one of the fundamental irreps of su(3).

\section{Conclusion \label{sec:4}}

The electronic structure of benzene and borazine was studied with the tight-binding formalism. Nearest neighbours dominate the shape of the electronic levels with the appropriate values of evanescence lengths for all type of atoms. Theoretical models with dominating second-neighbour couplings were explored. The resulting spectra turned out to be asymmetric and in the extreme situation where the evanescence length is comparable to the bond length between nearest neighbours, a triplet state was found or doublet- singlet inversions could be produced. The discrete version of the continuity equation and discrete current descriptions of probability were obtained. The particular case of nearest neighbours was computed and depicted, suporting the aromaticity interpretation for both C$_6$H$_6$ and B$_3$N$_3$H$_6$ compounds. It was confirmed that the introduction of a magnetic flux controls the chirality of the vorticity. The $C_3$ symmetry breaking  was explored by the introduction of defects in on-site energies and couplings that led to level splittings, either linear or quadratic, respectively. This might be important in the molecular case, when the position of an atom may vary due to vibrational motion (regarded as a well-known mechanism of electronic de-excitation) or when the substitution H $\rightarrow$ F in fluorination takes place. With algebraic techniques, we were able to extract the most salient features of homogeneous and composite hexagonal rings. Moreover, the existence of a supersymmetry was found, explaining the up-down symmetry of the energy levels an, in more refined cases, a sum rule of energy midpoints. 

Interestingly, the generality of our results can be used to propose the construction of artificial realizations in microwaves \cite{Kuhl_2010,Barkhofen_2013,Villafane_2013,Bellec_2013} and elastic vibrations \cite{Torrent_2012,Torrent_2013,Zhong_2011}. The latter are suported by recent engineering of evanescent couplings in solids \cite{Ramirez_Ramirez_2018, flores_2001, ramirez_2019}, much in the spirit of quantum tunneling.

\section*{References}
\bibliographystyle{iopart-num}


\begin{thebibliography}{10}
\expandafter\ifx\csname url\endcsname\relax
  \def\url#1{{\tt #1}}\fi
\expandafter\ifx\csname urlprefix\endcsname\relax\def\urlprefix{URL }\fi
\providecommand{\eprint}[2][]{\url{#2}}

\bibitem{cooper_electronic_1986}
Cooper D~L, Gerratt J and Raimondi M 1986 {\em Nature\/} {\bf 323} 699--701

\bibitem{benfey_august_1958}
Benfey O~T 1958 {\em J. Chem. Educ.\/} {\bf 35} 21

\bibitem{armstrong_electronic_1970}
Armstrong D~R and Clark D~T 1970 {\em J. Chem. Soc. D\/}  99



\bibitem{Parker_2018}
Parker S~F 2018 {\em {RSC} Advances\/} {\bf 8} 23875--23880

\bibitem{Kiran_2001}
Kiran B, Phukan A~K and Jemmis E~D 2001 {\em Inorganic Chemistry\/} {\bf 40}
  3615--3618

\bibitem{Shen_2007}
Shen W, Li M, Li Y and Wang S 2007 {\em Inorganica Chimica Acta\/} {\bf 360}
  619--624

\bibitem{Clark_1967}
Clark I~D and Frost D~C 1967 {\em Journal of the American Chemical Society\/}
  {\bf 89} 244--247

\bibitem{Doering_1977}
Doering J~P 1977 {\em The Journal of Chemical Physics\/} {\bf 67} 4065

\bibitem{Moskowitz_1963}
Moskowitz J~W and Barnett M~P 1963 {\em The Journal of Chemical Physics\/} {\bf
  39} 1557--1560

\bibitem{Kac_1977}
Kac V 1977 {\em Advances in Mathematics\/} {\bf 26} 8--96

\bibitem{Haag_1975}
Haag R, {\L}opusza{\'{n}}ski J~T and Sohnius M 1975 {\em Nuclear Physics B\/}
  {\bf 88} 257--274

\bibitem{Wootters_1987}
Wootters W~K 1987 {\em Annals of Physics\/} {\bf 176} 1--21

\bibitem{Vourdas_2006}
Vourdas A 2006 {\em Journal of Mathematical Physics\/} {\bf 47} 092104

\bibitem{Vourdas_2005}
Vourdas A 2005 {\em Journal of Physics A: Mathematical and General\/} {\bf 38}
  8453--8471

\bibitem{Vourdas_1993} Vourdas A and Bendjaballah C 1993 {\em Phys. Rev.A \/} {\bf 47} 3523

\bibitem{Zak_2005} Mann A, Revzen M and Zak J 2005 {\em J. Phys. A: Math. Gen.\/} {\bf 38} L389

\bibitem{Sadurni_2019}
Sadurn{\'{\i}} E and Hern{\'{a}}ndez-Espinosa Y 2019 {\em J. Phys. A: Math. and
  Theor.\/} {\bf 52} 295204




\bibitem{sadurni_diracmoshinsky_2011}
Sadurn\'i E 2011 {\em AIP Conf. Proc.\/} {\bf 1334} 249



\bibitem{stegmann_vortices_2019}
Stegmann T, Franco-Villafa\~ne J~A, Ortiz Y~P, Deffner M, Herrmann C, Kuhl U,
  Mortessagne F, Leyvraz F and Seligman T~H 2019 {\em ChemRxiv\/} {\bf
  Preprint}



\bibitem{Kuhl_2010}
Kuhl U, Barkhofen S, Tudorovskiy T, St\"ockmann H~J, Hossain T, de~Forges~de
  Parny L and Mortessagne F 2010 {\em Phys. Rev. B\/} {\bf 82} 094308

\bibitem{Barkhofen_2013}
Barkhofen S, Bellec M, Kuhl U and Mortessagne F 2013 {\em Phys. Rev. B\/} {\bf
  87} 035101

\bibitem{Villafane_2013}
Franco-Villafa{\~{n}}e J~A, Sadurn{\'{\i}} E, Barkhofen S, Kuhl U, Mortessagne
  F and Seligman T~H 2013 {\em Phys. Rev. Lett.\/} {\bf 111} 170405

\bibitem{Bellec_2013}
Bellec M, Kuhl U, Montambaux G and Mortessagne F 2013 {\em Phys. Rev. B\/} {\bf
  88} 115437

\bibitem{Torrent_2012}
Torrent D and S{\'{a}}nchez-Dehesa J 2012 {\em Phys. Rev. Letters\/} {\bf 108}
  174301

\bibitem{Torrent_2013}
Torrent D, Mayou D and S{\'{a}}nchez-Dehesa J 2013 {\em Phys. Rev. B\/} {\bf
  87} 115143

\bibitem{Zhong_2011}
Zhong W and Zhang X 2011 {\em Phys. Lett. A\/} {\bf 375} 3533--3536

\bibitem{pauling_shared-electron_1928}
Pauling L 1928 {\em Proc. Natl. Acad. Sci. USA\/} {\bf 14} 359

\bibitem{empedocles_electronic_1964}
Empedocles P~B and Linnett J~W 1964 {\em Proc. Royal Soc. Lond. Ser. A\/} {\bf
  282} 166--177

\bibitem{Gauss_2000}
Gauss J and Stanton J~F 2000 {\em J. Phys. Chem. A\/} {\bf 104} 2865--2868





\bibitem{Harrison_1989}
Harrison W~A 1989 {\em Electronic structure and the properties of solids: the
  physics of the chemical bond\/} ({DOVER} {PUBLICATIONS})

\bibitem{Zhao_2010}
Zhao K, Zhao M, Wang Z and Fan Y 2010 {\em Physica E: Low-dimensional Systems
  and Nanostructures\/} {\bf 43} 440--445

\bibitem{Geim_2007}
Geim A~K and Novoselov K~S 2007 {\em Nature Materials\/} {\bf 6} 183--191

\bibitem{Lindholm_1967} Lindholm E and Jonsson B.-\"O 1967 {\em Chem. Phys. Lett. \/} {\bf 1} 501--503

\bibitem{Tegeler_1980} Tegeler E, Wiech G and Faessler A 1980 {\em J. Phys. B: Atom. Mol. Phys.} {\bf 13} 4771






\bibitem{engelberts_electronic_2005}
Engelberts J~J, Havenith R~W~A, van Lenthe J~H, Jenneskens L~W and Fowler P~W
  2005 {\em Inorg. Chem.\/} {\bf 44} 5266--5272

\bibitem{Harshbarger_1969}
Harshbarger W, Lee G~H, Porter R~F and Bauer S~H 1969 {\em Inorg. Chem.\/} {\bf
  8} 1683--1689

\bibitem{Watanabe_2004}
Watanabe K, Taniguchi T and Kanda H 2004 {\em Nature Materials\/} {\bf 3}
  404--409


\bibitem{Castro_Neto_2009}
Neto A~H~C, Guinea F, Peres N~M~R, Novoselov K~S and Geim A~K 2009 {\em Rev.
  Mod. Phys.\/} {\bf 81} 109--162


\bibitem{Slater_1930}
Slater J~C 1930 {\em Phys. Rev.\/} {\bf 36} 57--64

\bibitem{Snyder_1971}
Snyder L~C 1971 {\em J. Chem. Phys.\/} {\bf 55} 95--99

\bibitem{castanos_soluble_1991}
Casta\~nos O, Frank A, L\'opez R and Urrutia L~F 1991 {\em Phys. Rev. D\/} {\bf
  43} 544--547



\bibitem{Ramirez_Ramirez_2018}
Ram\'irez--Ram\'irez F, M\'endez--S\'aanchez R~A, Baez G, Morales A, Guti\'errez L and
  Flores J 2018 Emulating tunneling with elastic vibrating beams {\em 2018
  Progress in Electromagnetics Research Symposium ({PIERS}-Toyama)\/} ({IEEE})
 


\bibitem{flores_2001} Morales A, Flores J, Guti\'errez L and M\'endez--S\'anchez R A 2002 {\em J. Acoust. Soc. Am. \/} {\bf 112} 1961--1967

\bibitem{ramirez_2019} Ram\'irez--Ram\'irez F, Flores--Olmedo E, B\'aez G, Sadurn\'i E and M\'endez--S\'anchez R A 2019 {\em arxiv1911.11272 \/} {\bf Preprint}




\end{thebibliography}

\providecommand{\newblock}{}

\end{document}